\documentclass[twocolumn]{aastex701}

\usepackage{amsmath}

\usepackage{graphicx}
\newcommand{\uhm}{Department of Physics and Astronomy, University of Hawai`i at M\=anoa, 2505 Correa Rd., Honolulu, HI, 96822, USA}
\begin{document}

\title{Calibrating Mid-Infrared Emission Features As Diagnostics of Star Formation in Infrared-Luminous Galaxies via Radiative Transfer Modeling}

\author[0000-0003-0796-362X]{Loren Robinson}
\affiliation{Department of Astronomy, University of Wisconsin-Madison, 475 N. Charter St., Madison, WI 53706, USA}
\email{ljrobinson4@wisc.edu}

\author[0000-0003-1748-2010]{Duncan Farrah}
\affiliation{\uhm}
\affiliation{Institute for Astronomy, University of Hawai‘i, 2680 Woodlawn Dr., Honolulu, HI 96822, USA}
\email{dfarrah@hawaii.edu}

\author[0000-0002-2612-4840]{Andreas Efstathiou}
\affiliation{School of Sciences, European University Cyprus, Diogenes Street, Engomi, 1516 Nicosia, Cyprus}
\email{A.Efstathiou@euc.ac.cy}

\author[0000-0001-6970-7782]{Athena Engholm}
\affiliation{\uhm}
\affiliation{Institute for Astronomy, University of Hawai‘i, 2680 Woodlawn Dr., Honolulu, HI 96822, USA}
\email{ae39@hawaii.edu}

\author[0000-0003-0917-9636]{Evanthia Hatziminaoglou}
\affiliation{European Southern Observatory, Karl-Schwarzschild-Str. 2, 85748 Garching bei München, Germany}
\affiliation{Instituto de Astrofisica de Canarias (IAC), E-38205 La Laguna, Tenerife, Spain}
\affiliation{Universidad de La Laguna, Opto. Astrofisica, E-38206 La Laguna, Tenerife, Spain}
\email{ehatzimi@eso.org}

\author[0000-0002-2877-0002]{Maya Joyce}
\affiliation{Michigan State University,Physics$\&$Astronomy Department, East Lansing, Michigan
 USA}
\email{joycemay@msu.edu}
 
\author[0000-0002-7716-6223]{Vianney Lebouteiller}
\affiliation{ Université Paris-Saclay, Université Paris Cité, CEA, CNRS, AIM, 91191, Gif-sur-Yvette, France}
\email{vianney.lebouteiller@cea.fr}

\author[0000-0003-0624-3276]{Sara Petty}
\affiliation{NorthWest Research Associates, 3380 Mitchell Ln., Boulder, CO 80301, USA}
\email{spetty@nwra.com}

\author[0000-0002-5206-5880]{L. Kate Pitchford}
\affiliation{Department of Physics and Astronomy, Texas A\&M University, College Station, TX, USA}
\affiliation{George \& Cynthia Woods Mitchell Institute for Fundamental Physics and Astronomy, Texas A\&M University, College Station, TX, USA}
\email{kpitchford@tamu.edu}

\author[0000-0002-9149-2973]{Jos\'e Afonso}
\affiliation{Instituto de Astrof\'{i}sica e Ci\^{e}ncias do Espa\c co, Universidade de Lisboa, Portugal}
\affiliation{Departamento de F\'{i}sica, Faculdade de Ci\^{e}ncias, Universidade de Lisboa, Portugal}
\email{jafonso@oal.ul.pt}

\author[0000-0002-9548-5033]{Dave L. Clements}
\affiliation{Imperial College London, Blackett Laboratory, Prince Consort Road, London, SW7 2AZ, UK}
\email{d.clements@imperial.ac.uk}

\author[0000-0002-3032-1783]{Mark Lacy}
\affiliation{National Radio Astronomy Observatory, Charlottesville, VA, USA}
\email{mlacy@nrao.edu}

\author[0000-0001-6139-649X]{Chris Pearson}
\affiliation{RAL Space, STFC Rutherford Appleton Laboratory, Didcot, Oxfordshire OX11 0QX, UK}
\affiliation{The Open University, Milton Keynes MK7 6AA, UK}
\affiliation{Oxford Astrophysics, University of Oxford, Keble Rd, Oxford OX1 3RH, UK}
\email{chris.pearson@stfc.ac.uk}

\author[0000-0001-6854-7545]{Dimitra Rigopoulou}
\affiliation{Oxford Astrophysics, University of Oxford, Keble Rd, Oxford OX1 3RH, UK}
\email{dimitra.rigopoulou@physics.ox.ac.uk}

\author{Michael Rowan-Robinson}
\affiliation{Imperial College London, Blackett Laboratory, Prince Consort Road, London, SW7 2AZ, UK}
\email{m.rrobinson@imperial.ac.uk}

\author[0000-0002-6736-9158]{Lingyu Wang}
\affiliation{ SRON Netherlands Institute for Space Research, Landleven 12, 9747 AD Groningen, The Netherlands}
\affiliation{Kapteyn Astronomical Institute, University of Groningen, Postbus 800, 9700 AV Groningen, The Netherlands}
\email{l.wang@sron.nl}

\begin{abstract}

Luminous infrared galaxies are key sites of obscured stellar mass assembly at $z>0.5$. Their star formation rates (SFRs) are often estimated using the luminosities of the $6.2\mu$m and $11.2\mu$m polycyclic aromatic hydrocarbon (PAH) features, or those of the [\ion{Ne}{2}] and [\ion{Ne}{3}] fine-structure lines, as they are minimally affected by obscuration. It is uncertain whether the calibration of these features as SFR tracers depends on the starburst bolometric luminosity or the level of Active Galactic Nucleus (AGN) activity. We here investigate the relationship between the luminosities of PAH and Neon lines with star formation rate for highly luminous objects using radiative transfer modeling and archival observations of 42 local Ultraluminous ($\ge10^{12}\mathrm{L}_\odot$) Infrared Galaxies (ULIRGs). We find that PAH and [\ion{Ne}{2}] features arise mainly in star-forming regions, with small contributions from the AGN or host, but that the [\ion{Ne}{3}] line has a mixed contribution from both star formation and AGN activity. We present relations between $\mathrm{L}_{\rm{PAH}}$ and $\mathrm{L}_{[\rm{NeII}]}$, and both starburst luminosity and SFR. We find relations for lower luminosity ($\mathrm{L}_{\rm{IR}}\simeq10^{10}-10^{12}\mathrm{L}_\odot$) systems underestimate the SFRs in local ULIRGs by up to $\sim$1 dex. The 6.2$\mu$m and 11.2$\mu$m PAH features, and the [\ion{Ne}{2}] line, are thus good tracers of SFR in ULIRGs. We do not find that a more luminous AGN affects the relationship between SFR and PAH or Neon luminosity, but that it can make PAH emission harder to discern. Our results and derived relations are relevant to studies of star-forming and composite galaxies at $z<3$ with the \textit{James Webb Space Telescope}.
\end{abstract}


\section{Introduction} \label{sec:intro}
\setcounter{footnote}{0}
Ultraluminous Infrared Galaxies (ULIRGs) are galaxies with infrared (IR; rest-frame $1 - 1000\,\mu$m) luminosities above $10^{12} \mathrm{L}_\odot$. They were first discovered in significant numbers by the Infrared Astronomical Satellite \citep[IRAS;][]{1984Soifer,1985Houck,Soifer1991-PropertiesofIRGalaxies}. The power source behind their IR emission is a combination of starburst and active galactic nucleus (AGN) activity, most of which is obscured \citep{1998Genzel, Rigopoulou1999, 2003Farrah}. ULIRGs are rare at $z < 0.3$, having a surface density of less than one per one hundred square degrees, but become more common at high redshift, with several hundred per square degree at $ z \ge 1$ \citep{Rowan-Robinson1997, Dole2001, Borys2003, Mortier2005, lefloch2005, Vaccari2010-HerMESSPIRE, Austermann2010, Goto2011}. Thus, high redshift ULIRGs play an important role in stellar mass assembly. Reviews of ULIRGs can be found in \citet{1996Sanders}, \citet{Lonsdal2006-ULIRG}, and \citet{Perez-Torres_Mattila_Alonso-Herrero_Aalto_Efstathiou_2021}.

The obscuration in ULIRGs makes estimating SFRs and supermassive black hole (SMBH) accretion rates challenging. Spectral features in the mid-infrared (MIR) are thus of great value, as the lower opacity of interstellar dust at these wavelengths means that the obscuration of these features is lower compared to those at ultraviolet (UV) through near-infrared wavelengths. Two groups of mid-infrared features are especially important for estimating SFRs. First are the polycyclic aromatic hydrocarbon (PAH) features at 3.3, 6.2, 7.7, 8.6, 11.2, and 12.5$\mu$m \citep[e.g.][]{Peeters2004}. These features result from the stretching and bending modes of PAH molecules, which are excited by UV and optical photons \citep{1984Leger_Puget, 1985Allamandola_Tielens_Barker, 1989Allamandola_Tielens_Barker, 2005Tielens, 2008Tielens, Hernan-Caballero2020}.  PAH emission is associated with star-forming regions as PAH molecules are excited by the relatively hard radiation from young stars \citep{Rigopoulou1999, Desai2007, Hernan-Caballero2020}. The 7.7 and 8.6$\mu$m PAH features are more difficult to measure against the 9.7$\mu$m silicate feature, which itself can be prominent in IR luminous systems. Thus, the 6.2 and 11.2$\mu$m PAH features are most suited for SFR estimation \citep{Hernan-Caballero2020}, and there exist several calibrations between the luminosities of these PAH features and SFR \citep{Shipley2016, Maragkoudakis2018, Cortsen2019, Xie_Ho_2019, Mordini2021}. Second are the fine-structure lines of [\ion{Ne}{2}] 12.8$\mu$m ($^2P_{1/2}$$ \rightarrow $$^2P_{3/2}$) and [\ion{Ne}{3}] 15.5$\mu$m ($^3P_1 $$\rightarrow $$^3P_2$).  These two lines are bright and easy to measure. The ionization potentials of [\ion{Ne}{2}] and [\ion{Ne}{3}] (21.56 and 40.96 eV, respectively) are such that they can arise in star-forming regions, and, like PAHs, they are less sensitive (than optical/UV lines) to dust extinction. Therefore, both these lines have been proposed as SFR indicators \citep{2007Farrah, Ho_Keto_2007, Zhuang2019, Xie_Ho_2019}. 

The PAH and Neon SFR calibrations become increasingly important with the advent of the \textit{James Webb} Space Telescope (JWST)/Mid-Infrared Instrument (MIRI), which has allowed for sensitive observation of the $\approx$5-28$\mu$m range to be extended to high redshifts (z$\sim$1.5 for [\ion{Ne}{2}], [\ion{Ne}{3}], 11.2$\mu$m PAH, z$\sim$3 for 6.2$\mu$m PAH), where the space density of ULIRGs is highest. In this regime, extracting accurate bolometric starburst, host, and AGN luminosities via SED fitting becomes difficult due to source faintness and the difficulty in obtaining well-sampled SEDs. As such, having an accurate, locally derived relation between MIR features and the physical properties within these galaxies is essential. 

While several relations have been derived for lower luminosity ($L_\mathrm{IR} \lesssim 10^{12}\,\mathrm{L}_\odot$) galaxies \citep{Maragkoudakis2018, Cortsen2019, Xie_Ho_2019}, few calibrations have been developed between MIR features and SFRs in ULIRGs specifically \citep{2007Farrah}. It is thus unclear whether the calibration between the PAH or Neon line luminosities and SFR is universal, or if it depends on factors such as starburst luminosity or the level of AGN activity. There is, however, some evidence that these calibrations may not be universal.  Studies have noted that PAH emission may be weaker in high luminosity \citep{Desai2007, Hernan-Caballero2020, 2023GOALS} or low metallically \citep{Xie_Ho_2019} systems, and that the AGN could either produce or suppress PAH emission \citep{2023GOALS}.  It also remains unclear whether PAH emission arises predominantly in star-forming regions, or if significant emission can also arise in the host galaxy ISM \citep{Zhang_Ho_2023}.  Similarly, the narrow line region of AGN can produce [\ion{Ne}{2}] and [\ion{Ne}{3}], leading to a debate over the level of AGN contribution to these lines \citep{Zhang_Ho_2023}. 

Further potential uncertainty is introduced with the method used to map luminosities of spectral features to physical conditions in the galaxy. For example, energy balance approaches, e.g., MAGPHYS \citep{Cunha_Charlot_Elbaz_2008}, in which Star Formation Histories (SFH) are derived by balancing the energy absorbed by dust in the rest-frame UV with that re-radiated in the IR. These have the advantage of speed, but lack the details of radiative transfer. In contrast, a radiative transfer based approach, such as CYGNUS \citep{Efstathiou2022, 2024SMART} or \texttt{CLOUDY} \citep{2017CLOUDY}, models physical conditions in a system. This method is more physical but is rarely used due to computational cost.  

This paper explores the relationship between PAH and [\ion{Ne}{2}] and [\ion{Ne}{3}] emission features and SFR, AGN luminosity, and other characteristics of ULIRGs. We combine archival observations of these MIR emission features with the results of a radiative transfer modeling study \citep{Efstathiou2022}, which has separated the IR luminosities of ULIRGs into starburst, host, and AGN luminosities. We develop scaling relationships between component IR luminosities and observed PAH and Neon line luminosities in local ULIRGs. We review sample selection and methods in Section \ref{sec: method}. In Section \ref{sec: resdisc} we present and analyze our derived relations, as well as compare with previous works. Finally, we highlight conclusions in Section \ref{sec:conclusion}. Throughout, we assume \mbox{$H_0 = 70$\,km\,s$^{-1}$\,Mpc$^{-1}$}, \mbox{$\Omega = 1$}, and \mbox{$\Omega_{\Lambda} = 0.7$}.  We further assume $\mathrm{L}_\odot = 3.827 \cdot 10^{26}$W.

\section{Methods} \label{sec: method}

\subsection{Sample Selection}
Our sample includes the 42 ULIRGs observed by the HERschel ULIRG Reference Survey (HERUS, \citealt{Farrah2013-FIR, Spoon2013, Pearson2016, Clements2018}). The selection is detailed in \citet{Efstathiou2022}, and we briefly summarize it here.  Originally selected from the IRAS Faint Source Catalog \citep{1988_IRAS}, the sample includes all known ULIRGs with $z<0.3$ and an IRAS 60$\mu$m flux density greater than $\sim$2Jy. This sample is thus an almost unbiased subset of the low-redshift ULIRG population. 

Measurements of the luminosities and equivalent widths of the PAH features at 6.2$\mu$m and 11.2$\mu$m were obtained using the InfraRed Spectrograph \citep[IRS;][]{2004Houck} on the \emph{Spitzer} space telescope \citep{2004Werner}. Observations were taken as part of the guaranteed time observations program \citep{Desai2007}.  We eschewed the 7.7$\mu$m and 8.6$\mu$m features as the former can be challenging (more so than the 11.2$\mu$m feature) to measure in the presence of silicate absorption at 9.7$\mu$m, and the latter is faint \citep[]{2009Bernard-Salas}. The 6.2 and 11.2$\mu$m PAH features were obtained by integrating the flux above a spline interpolated continuum \citep{spoon22}. Out of 42 sources, 40 had luminosities for both the 6.2 and 11.2$\mu$m PAH features, and one (IRAS 13536+1836) had data only for the 11.2$\mu$m PAH luminosity. All 42 sources had data for both the [\ion{Ne}{2}] and [\ion{Ne}{3}] lines. The PAH and Neon line luminosities are presented in Table \ref{table: sample}.

\subsection{Physical properties}\label{sec: phys_params}
We obtain physical properties of our sample from \citet{Efstathiou2022}. This study performed Spectral Energy Distribution (SED) fitting to the archival ultraviolet through millimeter-wavelength data for our sample, using radiative transfer models for the starburst, the AGN, and the host galaxy. We adopt their results based on the CYGNUS AGN models as they provide the best overall fit to the SEDs. Thus, unlike previous works, we compare line luminosities to starburst, AGN, and host luminosities, and other physical parameters, as follows   \citep[see also tables 2 and 3 of][]{Efstathiou2022}:

\begin{itemize}
\item $L^{o}_{\rm Tot}$, $L^{c}_{\rm Tot}$: Total observed and anisotropy-corrected IR luminosity.

\item $L_{\rm Sb}$: IR luminosity of the starburst.

\item $L^{o}_{\rm AGN}$, $L^{c}_{\rm AGN}$ ($\mathrm{L}_\odot$): Observed and anisotropy-corrected IR luminosity of the AGN. The anisotropy correction arises from the axisymmetric structure of the AGN obscurer \citep{Efstathiou2022}.

\item $L_{\rm Host}$: IR luminosity of the host galaxy.

\item $\dot{M}_{\rm Sb}$ ($\mathrm{M}_\odot \mathrm{yr}^{-1}$):  Starburst SFR. The starburst SFR is averaged over the age of the starburst and does not include the host galaxy SFR.

\item $SB_{\rm age}$: Starburst age.

\end{itemize}

\subsection{Analysis}
As a non-parametric test for correlations between MIR line properties and physical properties of our sample, we adopt the Kendall-$\tau$ correlation coefficient. The correlation coefficient $\tau$ ranges from -1 to 1, where 

\begin{equation}
\tau = 
\left\{
    \begin{array}{lr}
        -1 & \text{strong anti-correlation}\\
         0 & \text{no correlation}\\
         1 & \text{strong correlation }
    \end{array}
\right\}
\end{equation}

\noindent The p-value is the fractional probability that the null hypothesis of no correlation cannot be rejected. We adopt $p < 0.05$ as evidence for a correlation. 

While the Kendall-$\tau$ tests tell us the likelihood of a correlation between MIR line properties and physical characteristics in our ULIRGs, they do not provide a conversion between these properties. As a parametric model, given that our sample is relatively small, we adopt a log-linear model:

\begin{equation}\label{eq:loglinscal}
 \mathcal{L}_{\rm Observable} = \alpha \cdot \mathcal{L}_{\rm Component} + \beta
\end{equation} 

\noindent where $\mathcal{L} \equiv \log(L)$ (\textit{L} in units of $\mathrm{L}_\odot$), $\mathcal{L}_{\rm Component}$ is the component luminosity (starburst, host, or AGN) and $\mathcal{L}_{\rm Observable}$ is the MIR-feature luminosity. Similarly, we define $\dot{\mathcal{M}}_{\rm Sb} \equiv \log(\dot{M}_{\rm Sb})$, and $\mathcal{W} \equiv \log(EW)$. The calibrations were derived via fitting using the Orthogonal Distance Regression \citep[ODR;][]{Boggs_Rogers_1990} algorithm.\footnote{As implemented within the Scipy ODRPACK Version 2.01}

\section{Results \& Discussion}\label{sec: resdisc}
We first consider the results from the Kendall-$\tau$ tests between the luminosities of the PAH and Neon lines and the component luminosities derived from the SED fits (\S\ref{sec: phys_params}). There is evidence for correlations between the luminosities of the emission features and at least one of starburst and host luminosity, but no evidence for any correlations with AGN luminosity  (Tables \ref{Table: pahkt} and \ref{table: nekt}).\footnote{To verify that the correlations are not a result of working in luminosity space, we here and in all subsequent cases also perform correlation tests in flux space. We find that the correlations did not weaken as a result of performing the test in flux space.}

This suggests an origin for these lines in star-forming regions and/or the quiescent ISM. To explore these connections further, we construct log-linear relations (Equation \ref{eq:loglinscal}) between the feature luminosities and these component luminosities. A summary of all relations can be found in Table \ref{table: pah_scaling_relations}.

\subsection{PAHs as Tracers of Star Formation}\label{ss:pahtrace}
Based on the correlations between  starburst luminosity and both 6.2 and 11.2$\mu$m PAH luminosities, we construct log-linear relations between these quantities:

\begin{equation}\label{eq: 62sb}
\mathcal{L}_{\rm 6.2} = (1.09 \pm 0.16)\mathcal{L}_{\rm Sb} - (3.88 \pm 1.92)
\end{equation}

\begin{equation} \label{eq: 11sb}
 \mathcal{L}_{\rm 11.2} = (0.87 \pm 0.14) \mathcal{L}_{\rm Sb} - (1.38 \pm 1.64)
\end{equation}

\begin{figure*}[ht!]
    \epsscale{1.1}
    \centering
    \plottwo{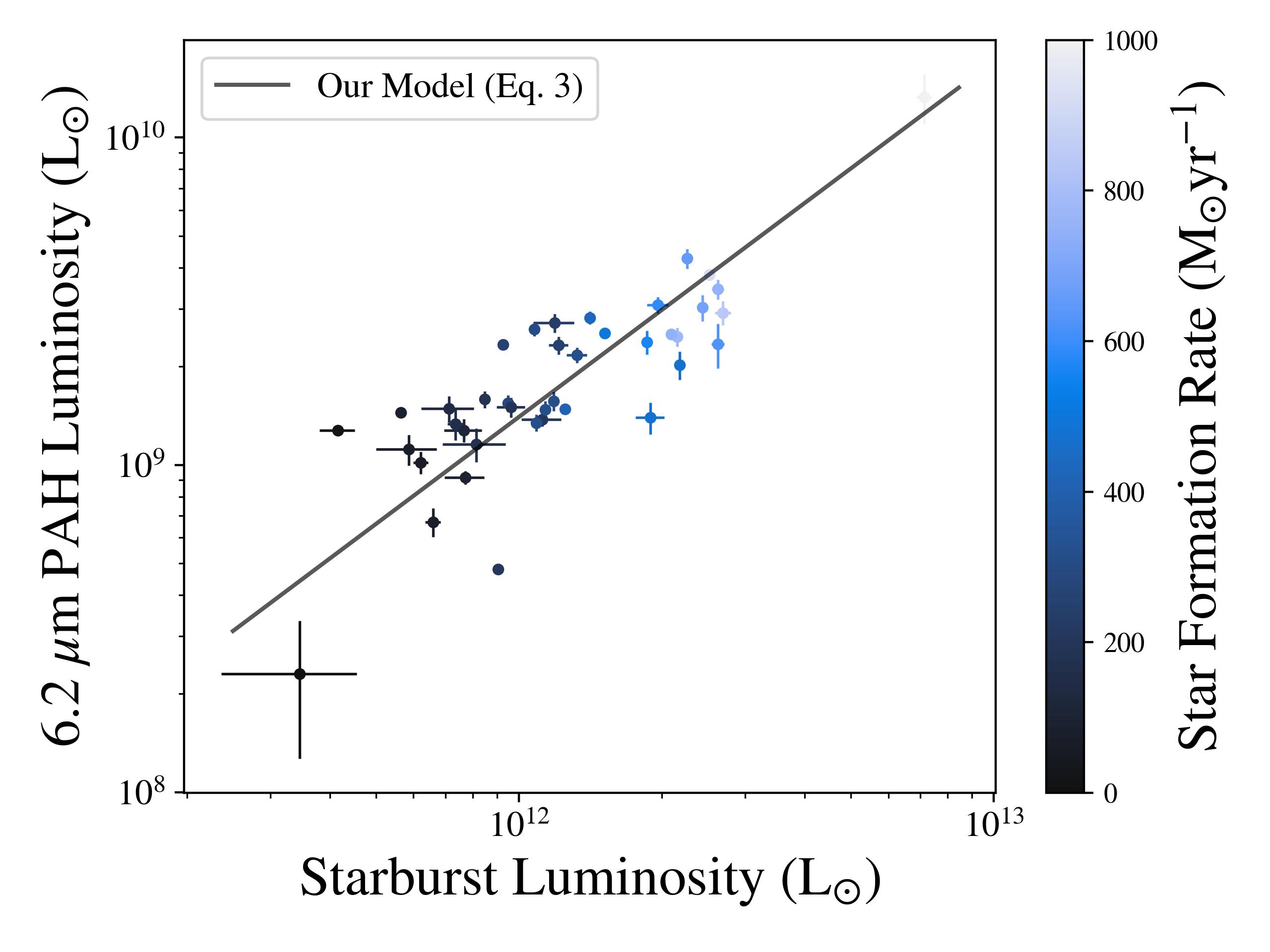}{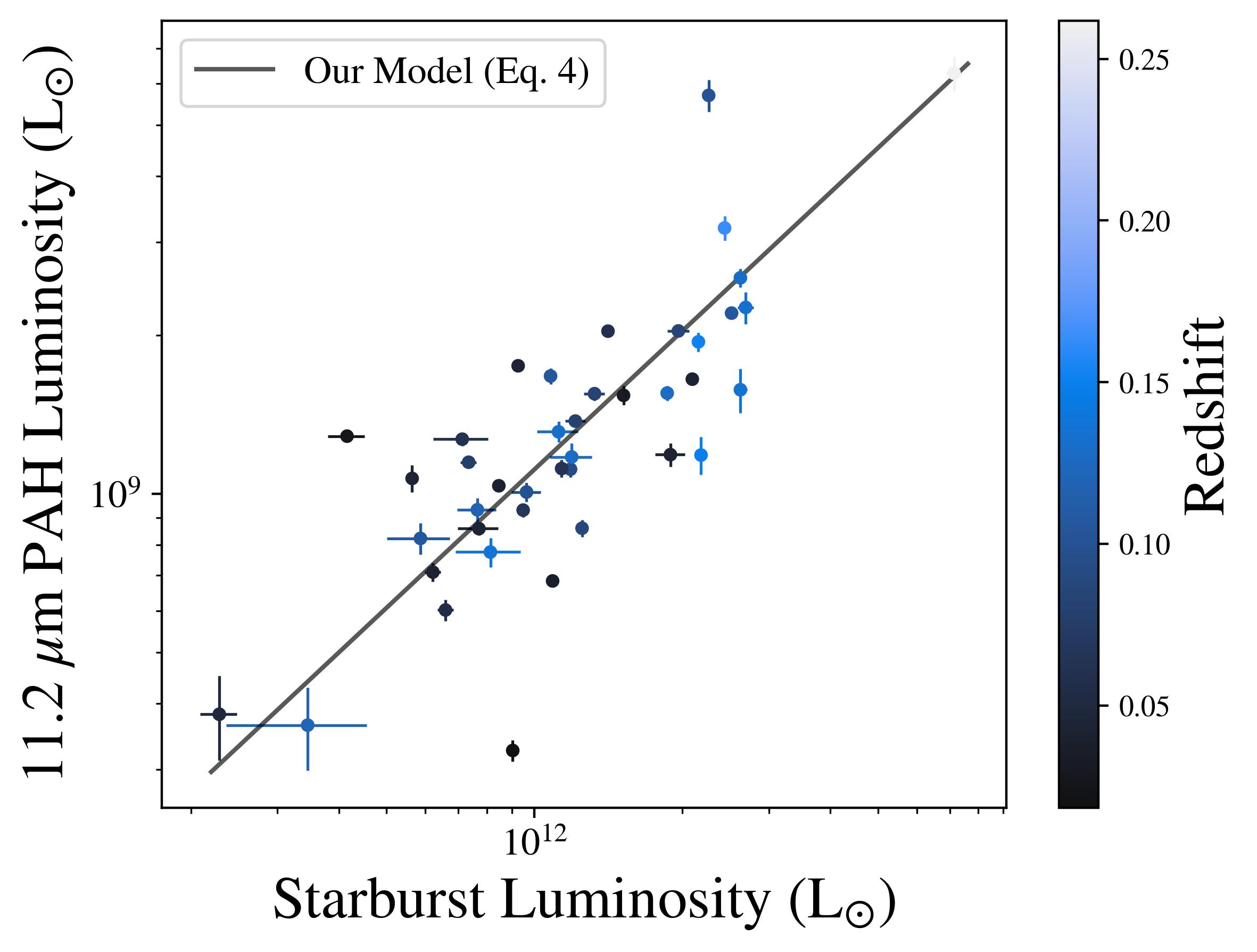}
    \caption{Starburst luminosity versus 6.2$\mu$m (left) and 11.2$\mu$m (right) PAH luminosity, color-coded by SFR (left) and redshift (right). Models derived above (Equations \ref{eq: 62sb} and \ref{eq: 11sb}) are plotted atop the data. 
    \label{fig:pahlsb_sfr}}
\end{figure*}

\noindent (Figure \ref{fig:pahlsb_sfr}). The slope of the 6.2$\mu$m relation is consistent with unity. A sub-unity slope is favored in the 11.2$\mu$m relation (though only at $1\sigma$ significance). Three factors could give a sub-unity slope. First, a more luminous starburst may be less efficient at producing PAH emission. Second, there may be greater dust dilution that suppresses PAH heating. Third, another factor may correlate with $L_{\rm Sb}$ and act to destroy PAHs or quench PAH emission. We discuss this possibility in \S\ref{sec: AGN_quenching}.

We next examine if our relations depend on any of the physical properties of the starburst (Table 2 from  \citealt{Efstathiou2022}) and the observed optical depth of the $9.7\mu$m silicate feature. To do so, we divide our sample into two subgroups with parameter values above/below the median and re-derive the relations. We only find evidence for a dependence in the relations with starburst age. The PAH relations for the subgroups of young ($SB_{\rm age}<30$Myr) and old ($SB_{\rm age}=30-35$Myr) starbursts are:

\begin{equation}
 \mathcal{L}_{\rm 6.2}^{\rm young} = (1.66 \pm 0.22) \mathcal{L}_{\rm Sb} - (11.02 \pm 2.64), 
\end{equation}

\begin{equation}
\mathcal{L}_{\rm 6.2}^{\rm old} = (0.51\pm 0.09) \mathcal{L}_{\rm Sb} + (3.16 \pm 1.05).
\end{equation}

\begin{equation}\label{eq: young11_lsb}
  \mathcal{L}_{\rm 11.2}^{\rm young} = (1.19 \pm 0.17) \mathcal{L}_{\rm Sb} - (5.31 \pm 2.12),  
\end{equation}

\begin{equation}\label{eq: old11_lsb}
  \mathcal{L}_{\rm 11.2}^{\rm old} = (0.67 \pm 0.19) \mathcal{L}_{\rm Sb} + (1.01 \pm 2.30)
\end{equation}

\begin{figure}[ht!] 
    \centering
    \includegraphics[width=\linewidth]{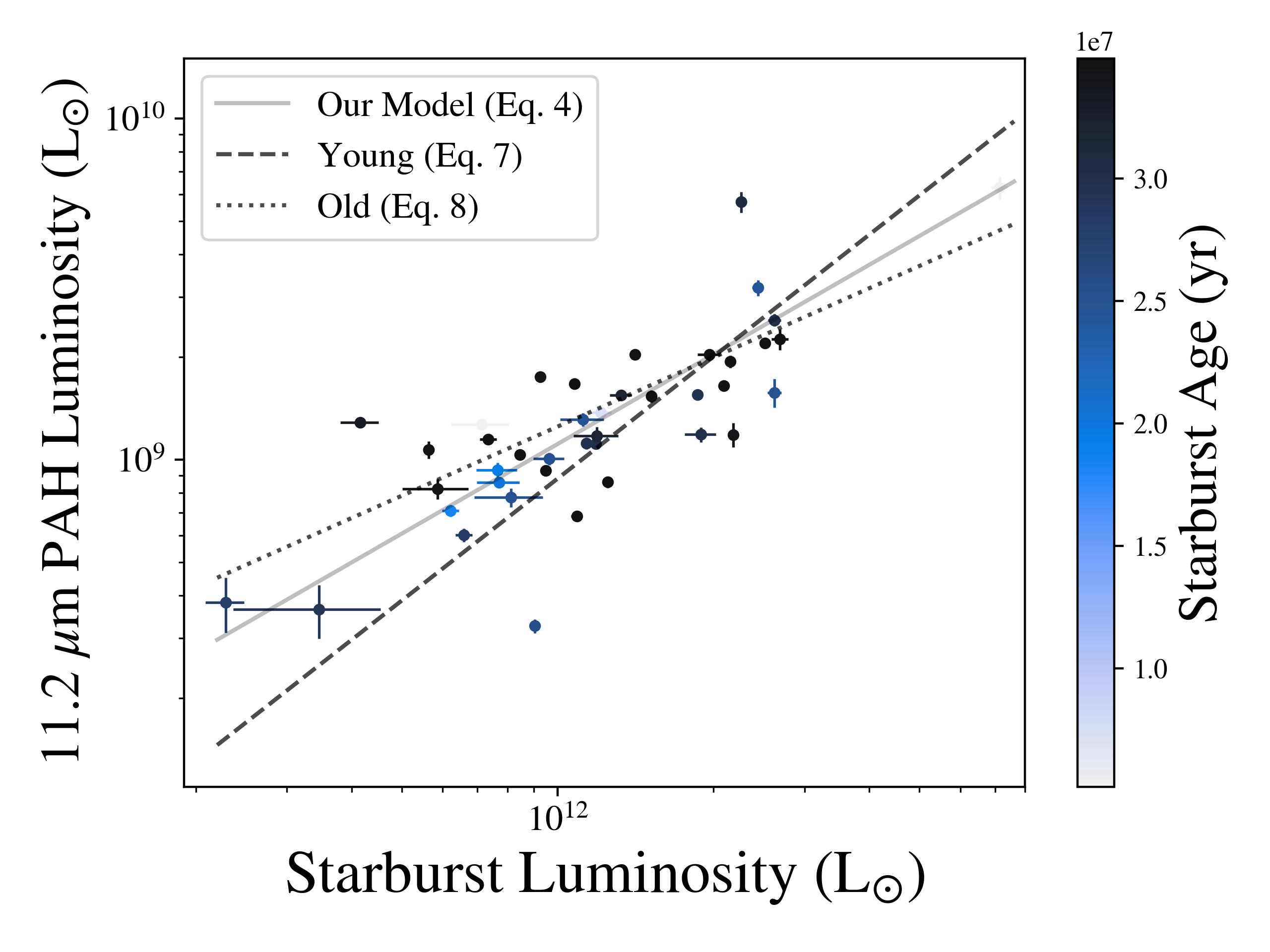}
    \caption{11.2$\mu$m PAH Luminosity versus $L_{\rm Sb}$, color-coded by starburst age. The dashed and dotted lines represent models derived for the young (Equation \ref{eq: young11_lsb}) and old (Equation \ref{eq: old11_lsb}) subgroups, respectively. An analogous plot of 6.2$\mu$m PAH Luminosity versus $L_{\rm Sb}$ is similar.
    \label{fig: oldyoung}}
\end{figure}

\noindent (Figure \ref{fig: oldyoung}). The shallower slopes of these relations for older starbursts are reasonable, as figure 3 in \citet{Efstathiou2000} shows that the PAH features become less intense with increasing age. These relations should, however, be used cautiously, as they are based on fits to only 20 objects. The slopes are, however, robust to varying the subgroups by up to five objects either way around the median. 

Finally, since starburst luminosity does not linearly translate to SFR in the \citet{Efstathiou2000} model, we derive scaling relations between PAH luminosity and SFR:

\begin{equation}\label{eq: 62sfr}
 \mathcal{L}_{\rm 6.2} = (0.95 \pm 0.12) \dot{\mathcal{M}}_{\rm Sb} + (6.81 \pm 0.31)
\end{equation}

\begin{equation}\label{eq: 11sfr}
\mathcal{L}_{\rm 11.2} = (1.08 \pm 0.16) \dot{\mathcal{M}}_{\rm Sb} + (6.32 \pm 0.42).
\end{equation}

\noindent As with starburst luminosity, we find that these SFR relations depend on starburst age - following the same procedure to divide the sample, we find:

\begin{equation}
\mathcal{L}_{\rm 6.2}^{\rm young} = (1.40 \pm 0.17) \dot{\mathcal{M}}_{\rm Sb} + (5.59 \pm 0.44),
\end{equation}

\begin{equation}
 \mathcal{L}_{\rm 6.2}^{\rm old} = (0.67 \pm 0.13) \dot{\mathcal{M}}_{\rm Sb} + (7.56 \pm 0.35).
\end{equation}

 \begin{equation}
 \mathcal{L}_{\rm 11.2}^{\rm young} = (1.45 \pm 0.23) \dot{\mathcal{M}}_{\rm Sb} + (5.41 \pm 0.58),
\end{equation}
 
\begin{equation}
 \mathcal{L}_{\rm 11.2}^{\rm old} = (1.08 \pm 0.26) \dot{\mathcal{M}}_{\rm Sb} + (6.32 \pm 0.71)
\end{equation}

\noindent for young and old starbursts, respectively. Again, older starbursts show statistically shallower slopes. We find no dependence on any other physical parameter from \citet{Efstathiou2022}.

\subsection{Starburst \& Host PAH Fits}
Motivated by the possibility that PAH emission could also arise in the quiescent ISM, heated by ambient starlight, we explore multi-component log-linear fits of the PAH luminosities against the starburst and host luminosities simultaneously.  This yields (see Figure \ref{fig: multivar}):

\begin{multline}
\mathcal{L}_{\rm 6.2} = (0.87 \pm 0.23) \mathcal{L}_{\rm Sb} \\ + (0.74 \pm 0.20 )\mathcal{L}_{\rm Host} - (9.26 \pm 3.56)
\end{multline}

\begin{multline}
\mathcal{L}_{\rm 11.2} = (0.74 \pm 0.29) \mathcal{L}_{\rm Sb} \\ + (1.09 \pm 0.40)\mathcal{L}_{\rm Host} -(11.68 \pm 5.24)
\end{multline}

\noindent These fits may suggest that the host galaxy contributes a small fraction of the PAH emission, as the slopes for $L_{\rm Host}$ are positive and comparable to the slope of $L_{\rm Sb}$. However, on average, $L_{\rm Sb}$ is $\sim$17 times greater than $L_{\rm Host}$. Thus, the starburst is likely to be the dominant source of PAH emission in most cases. 

To facilitate comparisons with works that compare PAH luminosities to total IR luminosity but for systems without a significant AGN, we derive scaling relations between PAH luminosities and $L_{\rm Sb + Host}$, yielding:

\begin{equation}\label{eq:comb62}
\mathcal{L}_{\rm 6.2} = (1.19 \pm 0.17) \mathcal{L}_{\rm Sb + Host} - (5.22 \pm 2.05),
\end{equation}

\begin{equation}\label{eq:combo11}
\mathcal{L}_{\rm 11.2} = (0.99 \pm 0.14) \mathcal{L}_{\rm Sb + Host} - (2.85 \pm 1.76). 
\end{equation}
 
\subsection{Comparison to Previous PAH Relations}
Several studies have derived relations to use PAHs as starburst tracers in galaxies \citep[e.g.,][]{1984Leger_Puget,2007Farrah,2013Yamada, Maragkoudakis2018, Cortsen2019, Hernan-Caballero2020}. Here we compare our relations to commonly-used relations from other works whose samples and methods resemble ours: \citet{Shipley2016, Maragkoudakis2018, Cortsen2019, Xie_Ho_2019, Mordini2021}. By comparing to works whose samples are primarily less luminous systems, we aim to investigate the possibility of a universal $L_{\rm IR}$-$L_{\rm PAH}$ relationship. We first briefly summarize the results of these studies. \citet{Shipley2016} present relations between PAH luminosity and SFR, as derived from the extinction-corrected H$\alpha$ line luminosity, for a sample of 105 galaxies with total IR luminosities of $10^{9} - 10^{12} \mathrm{L}_\odot$.  Their relations are: 

\begin{equation}
\dot{\mathcal{M}} = (0.96\pm 0.04)\mathcal{L}_{\rm 6.2} - (7.82 \pm 0.09),
\end{equation}

\begin{equation}
\dot{\mathcal{M}} = (1.06 \pm 0.03)\mathcal{L}_{\rm 11.2} - (8.55 \pm 0.08).
\end{equation}

\noindent \citet{Maragkoudakis2018} present an SFR to PAH luminosity relation based on 39 \ion{H}{2} regions within two galaxies (M33 and M83). Their SFRs are calibrated from the Spitzer MIPS 24$\mu$m emission.

\begin{equation}
\mathcal{L}_{\rm 6.2} = (1.43)\dot{\mathcal{M}}_{\rm Sb} + (9.29), 
\end{equation}	

\begin{equation}
\mathcal{L}_{\rm 11.2}  = (1.39)\dot{\mathcal{M}}_{\rm Sb} + (9.42).
\end{equation}

\noindent \citet{Cortsen2019} derive two relations between $6.2\mu$m PAH luminosity and total IR luminosity. First: 

\begin{equation}\label{eq:cortsf}
\mathcal{L}_{\rm 6.2} = (0.98 \pm 0.03) \mathcal{L}_{\rm IR} - (1.89 \pm 0.30), 
\end{equation}

\noindent for 31 star-forming galaxies at $z<0.36$, with $L_{\rm IR} \approx 10^{10}-10^{12} \mathrm{L}_\odot$ and four star-forming galaxies (SFGs) at z = $1.016-1.523$ with $L_{\rm IR} \approx 10^{11.86-12.66} \mathrm{L}_\odot$ for a total of 35 SFGs. Second:

\begin{equation}\label{eq:cortall}
\mathcal{L}_{\rm 6.2} = (0.81 \pm 0.03) \mathcal{L}_{\rm IR} - (0.04 \pm 0.29), 
\end{equation}

\noindent which was derived from these SFGs plus 5 AGN+Composite galaxies, 9 local ULIRGs, and 6 high-redshift sub-millimeter galaxies. \citet{Xie_Ho_2019} use a sample including 196 normal star-forming galaxies ($L_{\rm IR} \approx 10^9-10^{11}\mathrm{L}_\odot$), 13 ULIRGs ($L_{\rm IR} \approx 10^{12}-10^{12.9}\mathrm{L}_\odot$), and 17 blue compact dwarfs, for a total of 226 galaxies at $z<0.3$. Fitting against SFRs derived from extinction corrected $L_{\rm Ne II + Ne III}$, they find

\begin{equation}
\dot{\mathcal{M}}  = (1.05\pm 0.04) \mathcal{L}_{\rm 6.2} -  (7.84\pm 0.05), 
\end{equation}

\begin{equation}
\dot{\mathcal{M}} = (1.01\pm 0.04)\mathcal{L}_{\rm 11.2} - (7.88 \pm 0.04). 
\end{equation}
 
\noindent \citet{Mordini2021} use a sample comprised of $z \sim 0.027$ star-forming galaxies, AGN, observed in the IR from a variety of samples (see their table 1). Measuring SFR from the total IR luminosity, they find 

\begin{equation}
\dot{\mathcal{M}}  =  (0.76\pm 0.03)\mathcal{L}_{\rm 6.2} - (5.27 \pm 0.04)
\end{equation}

\begin{equation}
\dot{\mathcal{M}} = (0.73\pm 0.02)\mathcal{L}_{\rm 11.2} - (5.00 \pm 0.03)
\end{equation}

\noindent for a sample of 142 SFGs and 56/77 AGN for 6.2 PAH and 11.2 PAH, respectively.\footnote{While ULIRGs are originally excluded from the sample, \citet{Mordini2021} states that the inclusion of ULIRGs would not drastically change the slope of their relation.}

\begin{figure}[ht!] 
    \centering
    \includegraphics[width=\linewidth]{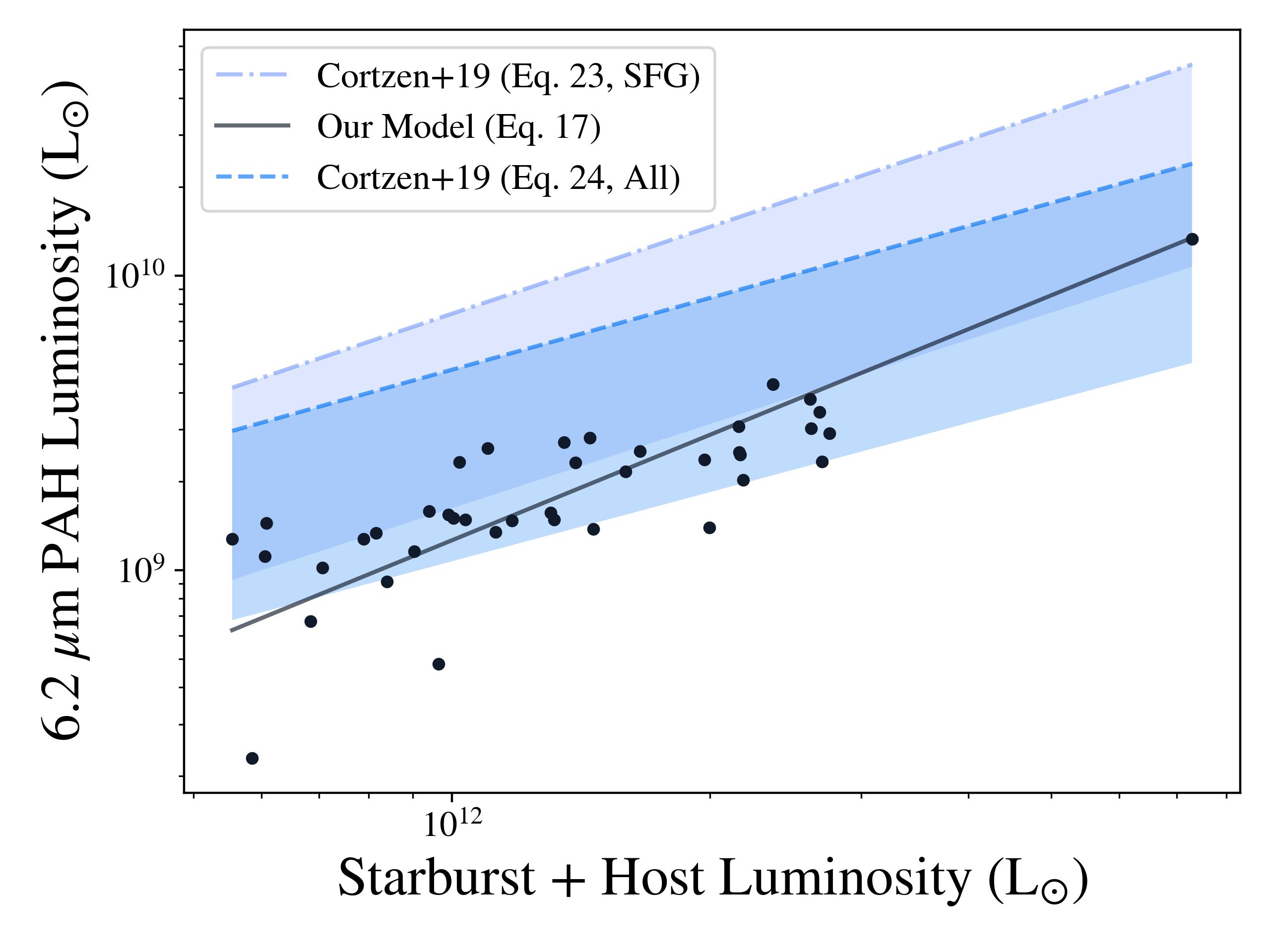}
    \caption{6.2$\mu$m PAH Luminosity versus $L_{\rm Sb} + L_{\rm Host}$, including models developed by \citet{Cortsen2019} (Equations \ref{eq:cortsf} and \ref{eq:cortall}, shaded to -1$\sigma$) and our own model (Equation \ref{eq:comb62}). We note that \citet{Cortsen2019} plots are against $L_{\rm Tot}$. However, their sample contained very few AGN sources, so $L_{\rm Sb} + L_{\rm Host}$ is the closest comparison.
    \label{fig:cortzen}}
\end{figure}

We first compare our relations to \citet{Cortsen2019} (Figure \ref{fig:cortzen}). Comparing our Equation \ref{eq:comb62} to Equation \ref{eq:cortsf}: the slopes and intercepts are consistent to $\pm 2 \sigma$. Instead comparing our Equation \ref{eq:comb62} to Equation  \ref{eq:cortall}: the slopes and intercepts are consistent to $\pm 3 \sigma$. At $L_{\rm Sb}+L_{\rm Host} = 1.315\times10^{12} \mathrm{L}_{\odot}$ Equations \ref{eq:cortsf} and \ref{eq:cortall} overestimate PAH luminosity in ULIRGs by 0.8 and 0.6\ dex, respectively.  This is consistent with previous works, which note that the ULIRGs have lower $L_{\rm PAH}/ L_{\rm IR}$ ratios \citep{Cortsen2019, Hernan-Caballero2020}.  This suggests that a universal $L_{\rm IR}$-$L_{\rm 6.2}$ relation across several dex cannot be established. Instead comparing with the relations from \citet{Maragkoudakis2018}, we find the slopes of our $6.2\mu$m PAH relations are inconsistent, while those of the $11.2\mu$m relations are within $3\sigma$. The intercepts of both relations are inconsistent.

\begin{figure*}[ht!]
    \epsscale{1.1}
    \centering
    \plottwo{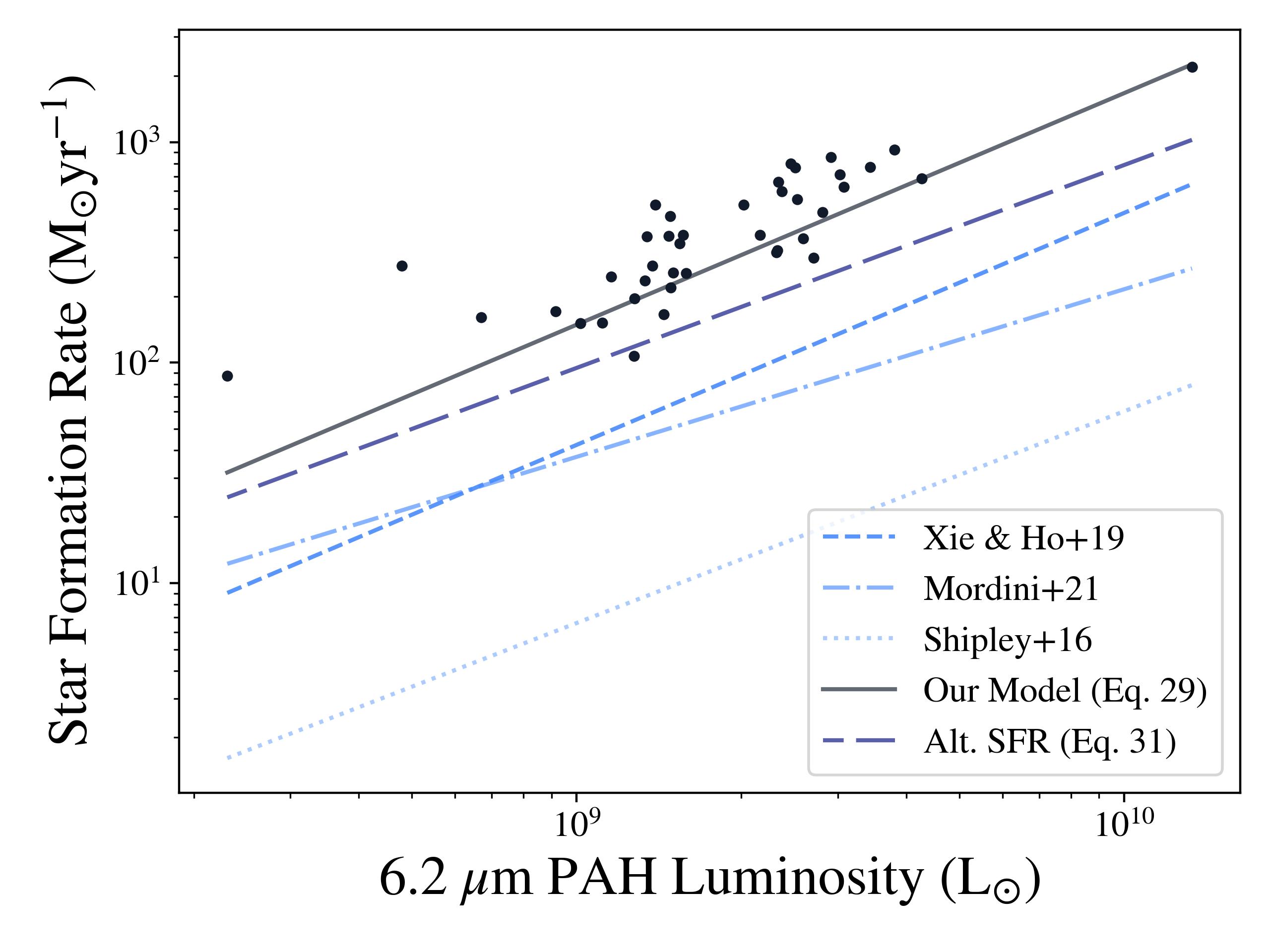}{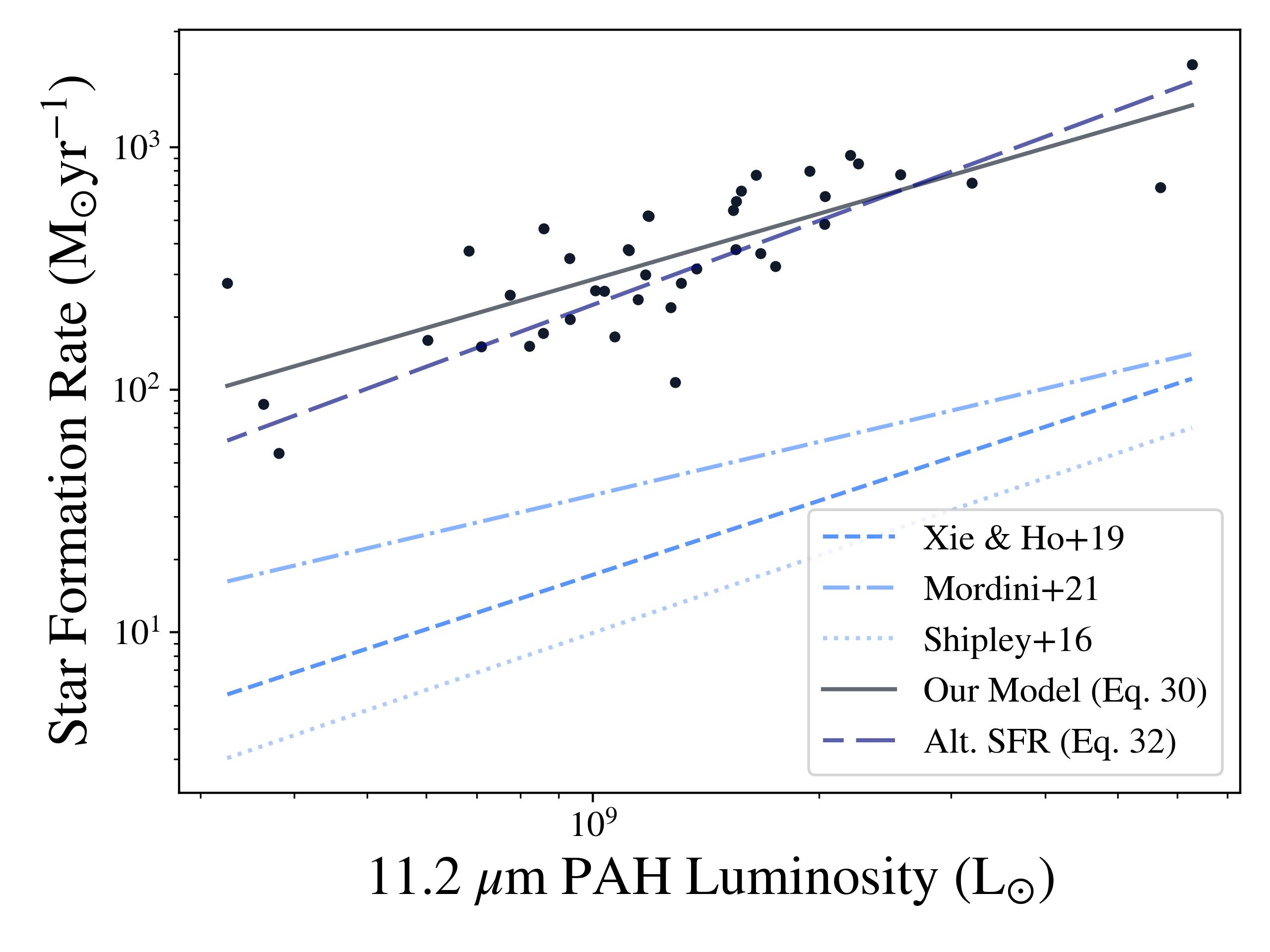}
    \caption{SFR versus both 6.2 (\textit{left}) and 11.2$\mu$m PAH luminosity (\textit{right}). Models derived by \citet{Shipley2016} (dotted line), \citet{Xie_Ho_2019} (dash line), and \citet{Mordini2021} (dash-dot line) are plotted atop the data, along with our own models (Equations \ref{eq: sfr62_alt} and \ref{eq: sfr11_alt}; black, solid line). We also include the model derived converting $L_{\rm Sb}$ to SFR (Equations \ref{eq: sfr62_kennicut} and \ref{eq: sfr11_kennicut}; long dash line).
    \label{fig: sfr_compare}}
\end{figure*}

The comparisons to \citet{Shipley2016}, \citet{Mordini2021}, and \citet{Xie_Ho_2019} require log-linear relations between PAH luminosity and SFR.  So, we re-derive our relations in this form, finding:

\begin{equation}\label{eq: sfr62_alt}
\dot{\mathcal{M}}  = (1.05\pm 0.13)\mathcal{L}_{\rm 6.2} - (7.28 \pm 5.50)
\end{equation}

\begin{equation}\label{eq: sfr11_alt}
\dot{\mathcal{M}}  = (0.93 \pm 0.13)\mathcal{L}_{\rm 11.2} - (5.79 \pm 5.75) 
\end{equation}

\noindent and compare to these three works in Figure \ref{fig: sfr_compare}.  Compared to the relations from \citet{Shipley2016}, our slopes for $6.2$ and $11.2\mu$m are consistent to $\pm 1\sigma,\pm 2\sigma $, respectively. Our relations imply that the relations in  \citet{Shipley2016} underestimate SFRs in ULIRGs by an average of 1.4 and 1.2 dex, for the $6.2$ and $11.2\mu$m features, respectively.  We note, though, that our SFRs are self-consistently calculated from the starburst model, while the \citet{Shipley2016} SFRs are calculated from H$\alpha$ line fluxes. While the H$\alpha$-based SFRs are extinction-corrected, the heavy obscuration in ULIRGs means that they may still underestimate the true SFR.   Compared to \citet{Mordini2021}, the slopes of our $6.2\mu$m and $11.2\mu$m relations are consistent to within $\sim 3\sigma, \sim 2\sigma$, respectively. Though the \citet{Mordini2021} relations underestimate the SFRs of our sample by an average of 0.7 dex. Compared to \citet{Xie_Ho_2019}, our $6.2$ and $11.2\mu$m slopes are consistent to $\sim 1\sigma$, but their relations underestimate our ULIRG SFRs by an average of 0.5 and 0.9 dex for the $6.2$ and $11.2\mu$m features, respectively.   To check if these differences arise from the differences in SFR calculation method mentioned above, we convert our starburst luminosity ($L_{\rm Sb}$) to a SFR following \citet{Kennicutt98}, and re-derive our relations. We obtain

\begin{equation}\label{eq: sfr62_kennicut}
\dot{\mathcal{M}}  = (0.92\pm 0.13)\mathcal{L}_{\rm 6.2} - (6.21 \pm 5.76)
\end{equation}

\begin{equation}\label{eq: sfr11_kennicut}
\dot{\mathcal{M}}  = (1.15 \pm 0.18)\mathcal{L}_{\rm 11.2} -  (8.20 \pm 7.67)
\end{equation}

\noindent Equation \ref{eq: sfr62_alt} differs from \citet{Shipley2016}, \citet{Mordini2021}, and \citet{Xie_Ho_2019} on average by 1.4, 0.7, and 0.5 dex, respectively. Equation \ref{eq: sfr62_kennicut} differs from those works by an average of 1.2, 0.5, and 0.4 dex, respectively. Similarly, Equation \ref{eq: sfr11_alt} differs from \citet{Shipley2016}, \citet{Mordini2021}, and \citet{Xie_Ho_2019} on average by 1.6, 1.0, and 1.3 dex, respectively. Equation \ref{eq: sfr11_kennicut} differs from the same relations by 1.2, 0.6, and 0.9 dex, respectively. The majority of the differences in the relations, therefore, must arise from origins other than the SFR calculation method. 

We ascribe the remaining differences in the relations to the generally higher luminosities of our sample compared to those used to derive the literature relations \citep[see also][]{2014Murata}. Broadly, there are two possible explanations for this. The first is that high luminosity starbursts are physically different from lower luminosity starbursts (or main sequence star formation).  For example, the mode of star formation in ULIRGs may be more compact than in less luminous star-forming galaxies \citep{2011Graci, 2017DiazSantos}. This could then mean that the lower luminosity starbursts have higher PAH heating efficiencies or more PAHs per unit stellar mass, as the more intense radiation field in the ULIRG starbursts could destroy PAHs.   It is also possible that higher obscuration in ULIRG starbursts means that some fraction of the PAH emission is extincted \citep{2014Murata}. If this were the case, we would expect more extinction in the 6.2$\mu$m feature than in the 11.2$\mu$m feature. A potential test of this is to examine whether the 11.2$\mu$m relations are closer to those in the literature than the 6.2$\mu$m relations. But differences in the features' origins, \citep[the 6.2$\mu$m feature originates from C–C stretching modes, while the 11.2$\mu$m feature originates in the C–H out-of-plane bending modes,][]{Peeters2004, Rigopoulou2021} mean that the 6.2$\mu$m feature will be much more prominent for ionized PAHs, while the converse is true for the 11.2$\mu$m feature. Thus, this is not a clean comparison, so we only note this as a possibility.  The second possibility is suppression of observed PAH emission by AGN activity, by processes such as the destruction of PAH molecules or the generation of a strong MIR continuum that drowns them out. We discuss this possibility in \S\ref{sec: AGN_quenching}.

\subsection{Neon Lines as Tracers of Star Formation}\label{ss: nesfr}
\begin{figure*}[ht!]
    \epsscale{1.1}
    \centering
    \plottwo{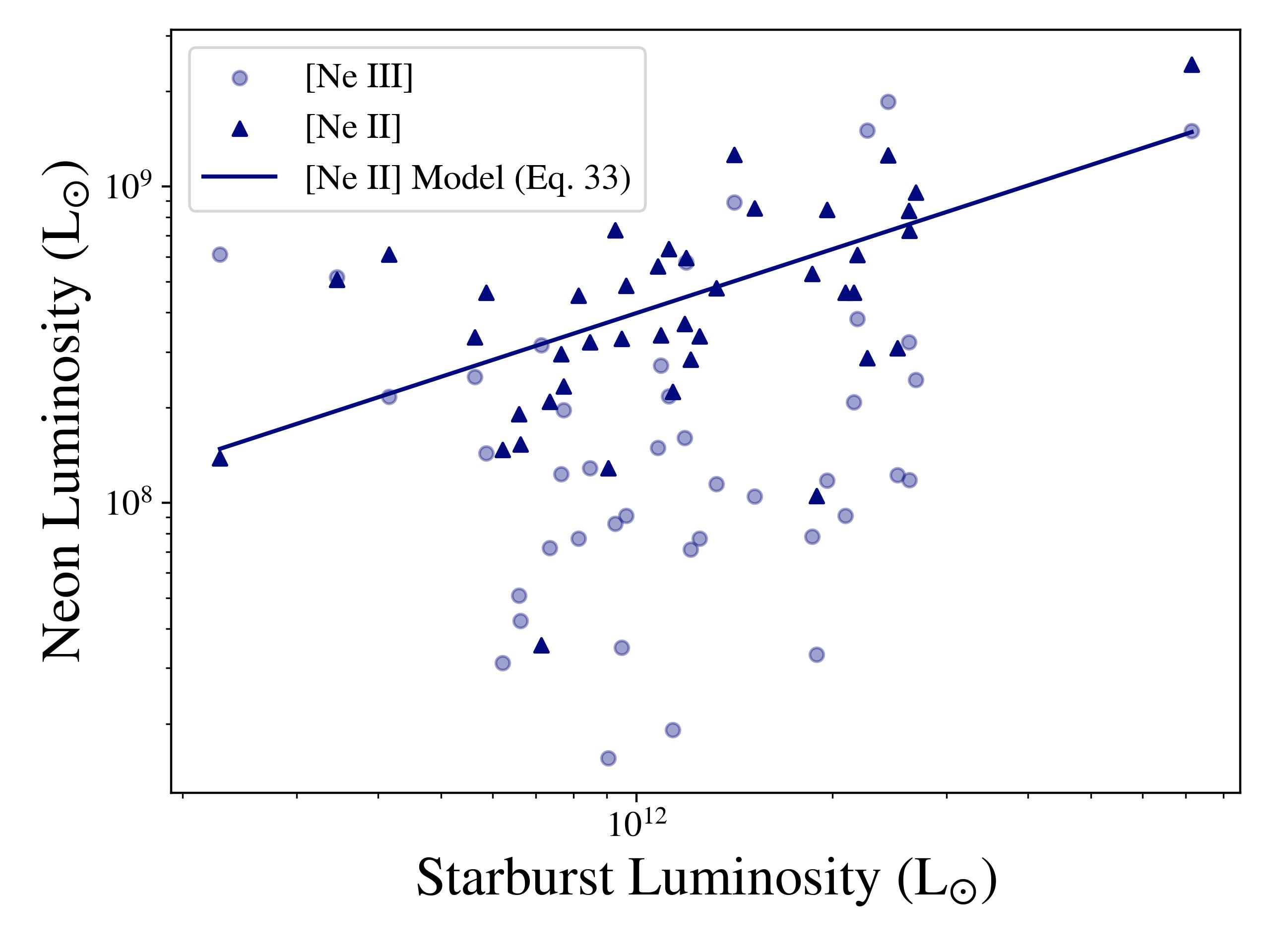}{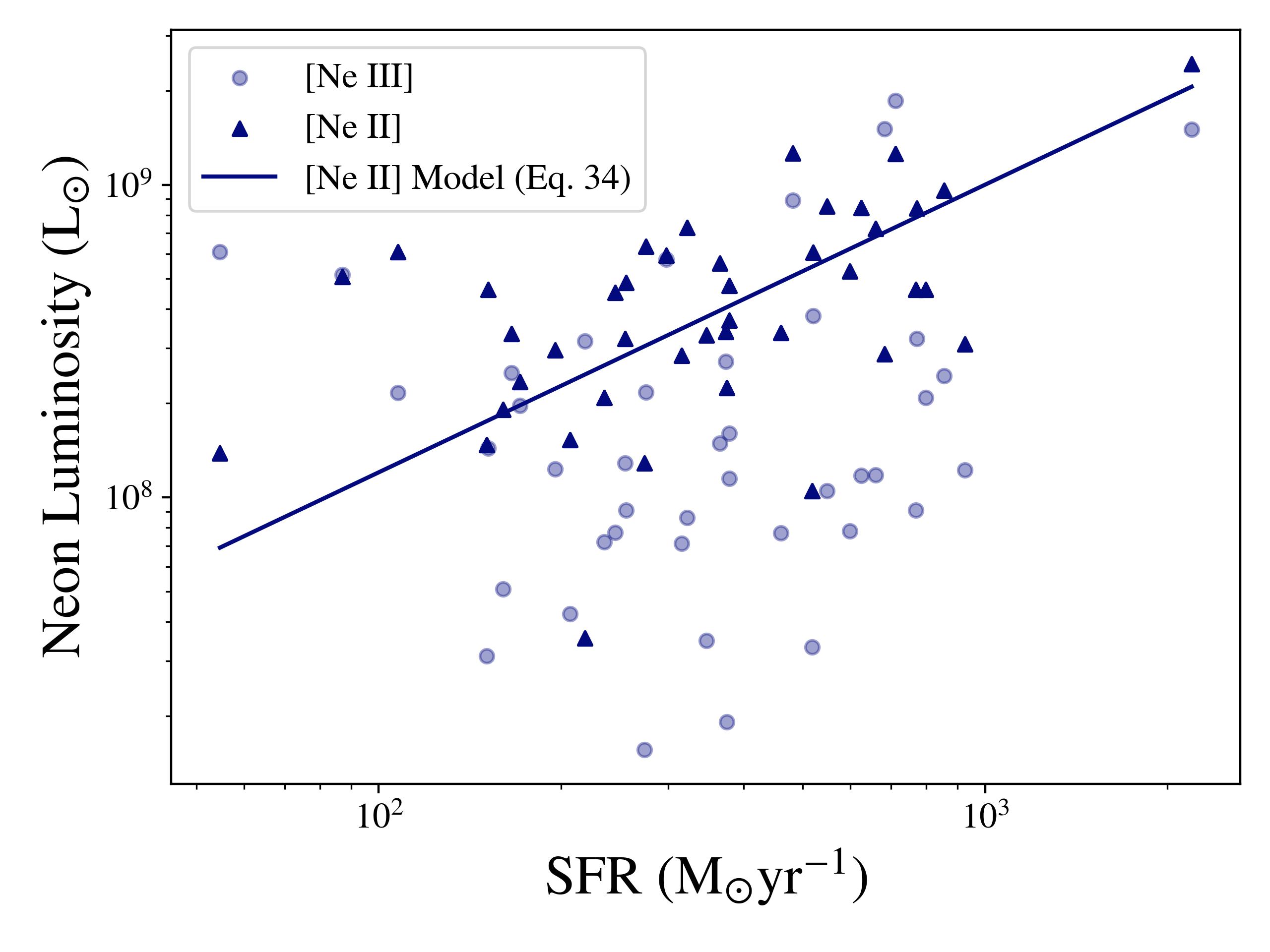}
     \caption{[\ion{Ne}{2}] (purple triangles) and  [\ion{Ne}{3}] (light purple circles) luminosity versus starburst luminosity (\textit{left}) and SFR (\textit{right}) with their corresponding models (Equations \ref{eq: neii_lsb} and \ref{eq: neii_sfr}) as solid lines. 
        \label{fig:ne_lsb}}
\end{figure*}

Using the same approaches as in \S\ref{ss:pahtrace}, we investigate relations between the Neon line luminosities and the luminosity of the starburst (Figure \ref{fig:ne_lsb}).   For the [\ion{Ne}{2}] $\lambda12.81\mu$m line luminosity, there is evidence for a correlation with $L_{\rm Sb}$ from the Kendall-$\tau$ tests. Fitting a log-linear relation yields:

\begin{equation}\label{eq: neii_lsb}
 \mathcal{L}_{\rm Ne II} = (0.67 \pm 0.09)\mathcal{L}_{\rm Sb} + (0.56 \pm 1.06).
\end{equation}

\noindent The equivalent relation with SFR is: 

\begin{equation}\label{eq: neii_sfr}
\mathcal{L}_{\rm Ne II} = (0.92 \pm 0.16) \dot{\mathcal{M}}_{\rm Sb} + (6.24 \pm 0.42).
\end{equation}

\noindent These relations are plotted atop the data in Figure \ref{fig:ne_lsb}. The $L_{\rm Ne II}$-$L_{\rm Sb}$ relation has a sub-unity slope at $\sim3\sigma$ significance, though the $L_{\rm Ne II}$-$\dot{M}_{\rm Sb}$ slope is consistent with unity at $< 1 \sigma$. We speculate that the closeness of the SFR slope to unity, in particular the possibility that it is closer to unity than the equivalent PAH relation, may signify that the [\ion{Ne}{2}] line is a better tracer of SFR than the brighter PAH features. 

For completeness, we also derive $L_{\rm NeII}$ to $L_{\rm Sb}$ and $L_{\rm Host}$, and $L_{\rm NeII}$ to $L_{\rm Sb + Host}$ relation. We find:

\begin{equation}
 \mathcal{L}_{\rm Ne II} = (0.73\pm 0.09)\mathcal{L}_{\rm Sb + Host} - (0.16 \pm 1.13)
\end{equation}

\begin{multline}
 \mathcal{L}_{\rm Ne II} = (0.65 \pm 0.18)\mathcal{L}_{\rm Sb} \\ + (0.56 \pm 0.16)\mathcal{L}_{\rm Host} - (5.28 \pm 2.57)
\end{multline}

\noindent respectively. 

Next, we search for dependence of the [\ion{Ne}{2}] line relations on the CYGNUS starburst properties, following the procedures detailed in \S\ref{ss:pahtrace}. Consistent with the analysis of the PAH features, we find evidence of a dependence on starburst age.  Younger starbursts have significantly steeper slopes than those derived for older starbursts. We find

\begin{equation}
 \mathcal{L}_{\rm Ne II}^{\rm young} = (1.10 \pm 0.14)\mathcal{L}_{\rm Sb} - (4.68 \pm 1.66)
\end{equation}
 
 \begin{equation}
 \mathcal{L}_{\rm Ne II}^{\rm old} = (0.53 \pm 0.08)\mathcal{L}_{\rm Sb} + (2.27 \pm 1.00)
\end{equation}

\noindent for young and old starbursts, respectively. For the SFR relations, we find

\begin{equation}
 \mathcal{L}_{\rm Ne II}^{\rm young} = (1.3 \pm 0.26 )\dot{\mathcal{M}}_{\rm Sb} + (5.25 \pm 0.66)
 \end{equation}
 
\begin{equation}
\mathcal{L}_{\rm Ne II}^{\rm old} = ( 0.69 \pm 0.21)\dot{\mathcal{M}}_{\rm Sb} + (6.87 \pm 0.56)
\end{equation}

\noindent for young and old starbursts, respectively.  It is plausible that this result signposts a dependency on SFR itself, since younger starbursts usually have higher SFRs than older starbursts. However, it is also possible that this signifies increased efficiency of Neon heating in younger starbursts. We find no evidence of dependence on the other CYGNUS starburst properties.

Turning to the [\ion{Ne}{3}] line: the evidence for a correlation between the [\ion{Ne}{3}] line and starburst luminosity is weaker. Examining the Kendall-$\tau$ results, we find no evidence for a correlation with either starburst or host luminosity. Deriving log-linear calibrations yields a nonzero slope, but the [\ion{Ne}{3}] luminosity in Figure \ref{fig:ne_lsb} has substantial scatter around the [\ion{Ne}{2}] relation, rendering these equations of limited value. A lack of correlation between [\ion{Ne}{3}] and star formation contradicts multiple studies that have found the [\ion{Ne}{3}] line to have a mixed contribution from both star formation and AGN activity \citep[e.g.,][]{Gorjian2007, Weaver2010, 2022Stone, Spinoglio2022}. AGN activity may interfere with our ability to detect a correlation between [\ion{Ne}{3}] and starburst luminosity. 

To investigate the extent to which [\ion{Ne}{3}] is affected by AGN activity, we conduct further analysis using the [\ion{Ne}{5}] lines at $14.32\mu$m and $24.32\mu$m. These lines are AGN tracers independent from SED modeling \citep[e.g.,][]{Gorjian2007, Mel'endez2008}. We divide the sample into two groups; one, the `detection' group, includes sources with detections of at least one of the two [\ion{Ne}{5}] lines, and the other includes objects with no detections of either [\ion{Ne}{5}] line. The sizes of these groups are 15 `detection' sources and 27 `non-detection' sources, respectively.  Using these groups, we repeat our analysis. We find evidence of correlations between [\ion{Ne}{2}] luminosity and starburst luminosity for both groups, with a stronger correlation in the non-detection group. This suggests that [\ion{Ne}{2}] emission comes primarily from star formation. For the non-detection group, we find a correlation between [\ion{Ne}{3}] luminosity and starburst luminosity. However, we find no correlation between [\ion{Ne}{3}] luminosity and starburst luminosity for the detection group.  We repeated this analysis using the sum of the [\ion{Ne}{2}] and [\ion{Ne}{3}] luminosities but found a correlation with starburst luminosity only in the non-detection group.  Moreover, the [\ion{Ne}{2}] luminosity shows strong positive correlations with both PAH luminosities, but the correlations between [\ion{Ne}{3}] luminosity and PAH luminosities are much weaker\footnote{This appears to contrast with \citet{2007Farrah}, who did not examine the Neon and PAH features separately, and found a correlation between  ($L_{\rm Ne II} + L_{\rm Ne III}$) and ($L_{\rm 6.2} + L_{\rm 11.2}$). We speculate that this difference is because we consider the Neon and PAH lines individually. To investigate this, we fit ($L_{\rm Ne II} + L_{\rm Ne III}$) to both starburst and host luminosities but find no evidence for an improvement in the significance of the relations. However, re-deriving their equation 5, we find consistency to within $3\sigma$.} (Table \ref{table: nekt}). Furthermore, we find a correlation between [\ion{Ne}{3}] and AGN luminosity in our detection group, but not in the non-detection group. 

We also examined if [\ion{Ne}{3}] luminosity is dependent on any CYGNUS starburst properties. In contrast to the results for PAH and [\ion{Ne}{2}] luminosity, we found negative correlations between  [\ion{Ne}{3}] luminosity and the starburst's initial optical depth and e-folding time.  We note, however, that the initial optical depth and e-folding time of starbursts are challenging to constrain tightly via SED modeling. 
 
From these analyses, we infer that [\ion{Ne}{2}] is mostly produced in star-forming regions, and can safely be used as a SFR indicator. However, [\ion{Ne}{3}] arises from a combination of starburst and AGN ionization, which leads to its relatively poor performance as a star formation tracer in composite objects \citep[see also][]{2009SpoonHolt}.  The large scatter in Figure \ref{fig:ne_lsb} is the result of significant AGN contribution to [\ion{Ne}{3}] emission, and possibly also variation in starburst properties.  Thus, [\ion{Ne}{3}] should only be used as a SFR indicator when there is little to no AGN presence, and even then with caution.

\subsection{Comparison to Previous Neon Relations}
Next, we compare  our $L_{\rm Sb}$ relation (Equation \ref{eq: neii_lsb}) to that of \citet{Ho_Keto_2007}. Their relation, derived for star-forming galaxies, between $L_{\rm Ne II}$ and $L_{\rm IR}$ is

\begin{equation}
  \mathcal{L}_{\rm Ne II } =  (1.01 \pm 0.05) \mathcal{L}_{\rm IR} - (3.44 \pm 0.56).   
\end{equation}

\begin{figure}[ht!]
    \centering
    \includegraphics[width=\linewidth]{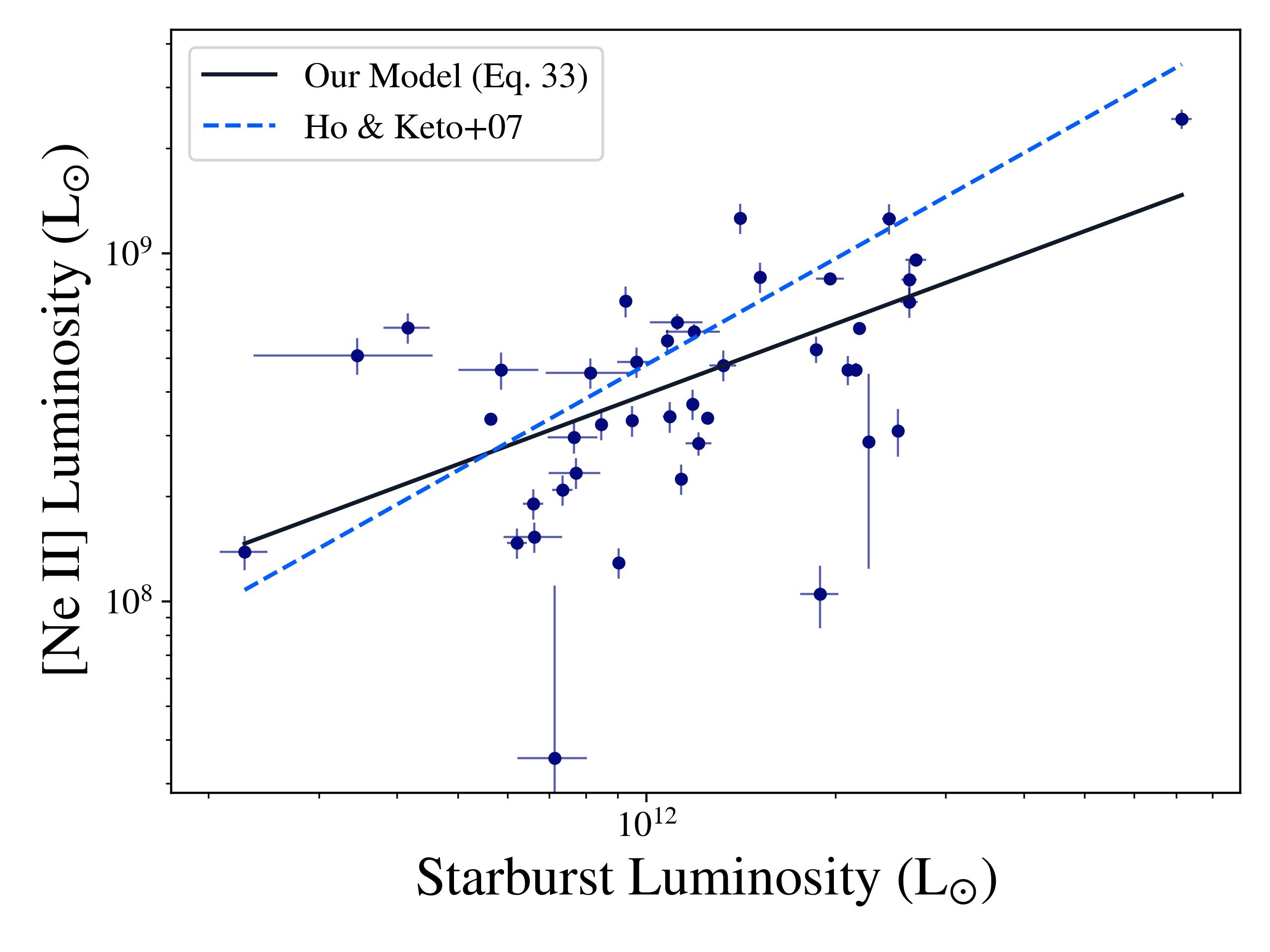}
    \caption{[\ion{Ne}{2}] luminosity versus $L_{\rm Sb}$, including model developed by \citet{Ho_Keto_2007} and our own model (Equation \ref{eq: neii_lsb}).
    \label{fig:ho_keto}}
\end{figure}

\noindent We plot these relations atop our sample in Figure \ref{fig:ho_keto}. Our relation has a shallower slope and a more positive intercept. However, unlike the PAH relations, our models predict more similar $L_{\rm Ne II}$. This is likely because \citet{Ho_Keto_2007} uses a sample of star-forming galaxies with luminosities spanning $10^7-\sim10^{12.5} \mathrm{L}_\odot$, the tail of the bright end of which overlaps with the luminosities of our sample ($\ge10^{12} \mathrm{L}_\odot$). They also note that the relation appears to steepen for less luminous systems ($L_{\rm IR}\le 10^{9.5} \mathrm{L}_\odot$).

\subsection{Effects of the AGN}\label{sec: AGN_quenching}
Our Kendall-$\tau$ tests found no evidence for a correlation between AGN luminosity and the luminosities of the PAH features and Neon lines. The possibility that AGN activity affects these star formation tracers has, however, been raised before, both observationally and theoretically \citep{Roche1991, Jensen2017, Cortsen2019, Hernan-Caballero2020, Rigopoulou2021}. So, we search for evidence that the AGN in our sample affect the relationships between the PAH \& Neon lines, and the starburst luminosities \& SFRs. To begin, we perform two basic tests.\footnote{We do not examine the correlation between the AGN luminosity and the $L_{\rm PAH} / L_{\rm IR}$ ratio, as a relatively more luminous AGN will cause this ratio to drop, without the AGN directly affecting the starbursts.}  First, we split the sample into two at $\frac{L_{\rm AGN}}{L_{\rm Tot}} = 0.2$ and re-derive the relations (Equations \ref{eq: 62sb} and \ref{eq: 11sb}); we find them to be statistically indistinguishable. This does not argue against AGN quenching of star formation, but does suggest that AGN do not substantially affect how starburst regions heat PAHs. Second, we perform the same test using the relations with [\ion{Ne}{2}] in Equations \ref{eq: neii_lsb}, \ref{eq: neii_sfr} and find no dependence. This suggests that AGN activity does not significantly impact [\ion{Ne}{2}] as an SFR tracer \citep{2022Stone}.

To perform a more nuanced test, we develop multi-variable relations between PAH luminosity and component luminosities.  Starting with the $6.2\mu$m PAH luminosity, we find the following relation using the observed AGN luminosity: 

\begin{multline}
\mathcal{L}_{\rm 6.2} = (1.05 \pm 0.17)\mathcal{L}_{\rm Sb} \\ - (0.48 \pm 0.11 )\mathcal{L}_{\rm AGN}^{o} + (2.03 \pm 1.74) 
\end{multline}

\noindent and

\begin{multline}
\mathcal{L}_{\rm 6.2} = ( 0.85 \pm 0.13) \mathcal{L}_{\rm Sb} \\- (0.30 \pm 0.07 )\mathcal{L}_{\rm AGN}^{c} + (2.51 \pm 1.43)  
\end{multline}

\noindent using the anisotropy-corrected AGN luminosity.  The equivalent relations using the PAH $11.2\mu$m luminosity are:

\begin{multline}
\mathcal{L}_{\rm 11.2} = (1.33 \pm 0.27) \mathcal{L}_{\rm Sb}\\  - (0.74 \pm 0.21)\mathcal{L}_{\rm AGN}^{o} + (1.52 \pm 2.47)  
\end{multline}

\begin{multline}
\mathcal{L}_{\rm 11.2} = (1.05 \pm 0.20) \mathcal{L}_{\rm Sb} \\ - (0.50 \pm 0.14)\mathcal{L}_{\rm AGN}^{c} + (2.30 \pm 2.17).  
\end{multline}

\noindent For the [\ion{Ne}{2}] line, the equivalent relations for the observed and anisotropy-correlated AGN luminosities are:

\begin{multline}\label{eq: ne2_lsb_agno}
\mathcal{L}_{\rm Ne II} = (1.16 \pm 0.20) \mathcal{L}_{\rm Sb} \\ - (0.42 \pm 0.14 )\mathcal{L}_{\rm AGN}^{o}- (0.47 \pm 1.89)
\end{multline}

\begin{multline}\label{eq: ne2_lsb_agnc}
\mathcal{L}_{\rm Ne II} = (0.97 \pm 0.16) \mathcal{L}_{\rm Sb} \\ - (0.41 \pm 0.11 )\mathcal{L}_{\rm AGN}^{c} +(1.87 \pm 1.76) 
\end{multline}

\noindent respectively.\footnote{Analogous equations for [\ion{Ne}{3}] showed slopes consistent with zero.} In all cases, we see weak but significant ($3-4\sigma$) negative slopes for the AGN component.  For the PAH relations, the slopes become less steep when the anisotropy-corrected AGN luminosity is used, though the significance of the slope being negative in general does not change much. This can be plausibly interpreted as negative AGN feedback, in which a more luminous AGN acts to suppress star formation but does not significantly affect how starburst regions produce PAH and Neon line emission. We cannot, however, exclude other possibilities such as, e.g., AGN phase lagging the starburst phase in time but having no effect on PAH emission; this scenario could give rise to a negative $L_{\rm AGN}$ slope in Equations \ref{eq: ne2_lsb_agno} and \ref{eq: ne2_lsb_agnc}. 

As a further test of AGN feedback, we examine how the EW of the PAH features depends on the component luminosities. The Kendall-$\tau$ test shows strong anti-correlations between the PAH EWs and AGN luminosity, but not with any other luminosity. This is consistent with the AGN luminosity alone determining the relative brightness of PAH features against the MIR continuum. We derive scaling relations between the PAH EWs and the AGN luminosity, finding: 

\begin{equation}
\mathcal{W}_{\rm 6.2}  = (-1.42 \pm 0.15) \mathcal{L}_{\rm AGN}^{o} + (15.51 \pm 1.68)    
\end{equation}

\begin{equation}
  \mathcal{W}_{\rm 11.2}  = (-1.95 \pm 0.33) \mathcal{L}_{\rm AGN}^{o} + (21.91 \pm 3.83)  
\end{equation}

\noindent for observed AGN luminosity, and:

\begin{equation}
\mathcal{W}_{6.2}  = (-1.96 \pm 0.47) \mathcal{L}_{\rm AGN}^{c} + (22.57 \pm 5.66) 
\end{equation}

\noindent for the anisotropy-corrected AGN luminosity.\footnote{We found no correlation between $\mathcal{L}_{\rm AGN}^{c}$ and $11.2\mu$m EW, and thus do not present a scaling relation between the two.} We do not find any dependence of these relations with PAH EW on any starburst, AGN, or host physical parameters, including starburst age.  A plausible interpretation is that a more luminous AGN produces a brighter MIR continuum, thus reducing the contrast of the PAH features \citep{Lutz1998, Desai2007, Alonso-Herrero2014, Jensen2017}. However, galaxies with greater AGN luminosity may have lower starburst luminosities, and therefore less PAH emissions \citep{Desai2007}. If this were the case, we would also observe an anti-correlation between $L_{\rm AGN}^{o}$ and $L_{\rm Sb}$, which we do not find. Our relations are consistent with the AGN being more adept at drowning out the 6.2$\mu$m PAH feature, consistent with the AGN SED peaking at MIR wavelengths \cite[see also e.g.][]{Jensen2017}. EW measurements for [\ion{Ne}{2}] and [\ion{Ne}{3}] have not been published, so we do not present a similar analysis for them.  

\section{Conclusions} \label{sec:conclusion}
Before we begin, it is important to note that the PAH and [\ion{Ne}{2}] relations above may not apply to objects with high IR luminosities at higher redshift ($1<z<3$). For instance, ``normal'' mode, main sequence galaxies at higher $z$ exhibit ULIRG-like luminosities \citep{Elbaz2011, Rujopakarn2011}, but have stronger PAH emission features compared to local ULIRGs \citep{Elbaz2011}. This is because their star formation mode is distinct from that of local ULIRGs (which are in starburst mode). Most IR luminous star-forming galaxies at high redshift are thought to have an \textit{extended} star-forming region, as opposed to the centralized region in local ULIRGs \citep{Rujopakarn2011, Rujopakarn2013, Bellocchi2022}. As such, the degree to which our relations hold depends on how the ISM phases from which PAHs and Neon lines arise are affected by the physical change in star-forming regions with redshift. We suggest that future works examine the efficacy of these relations on IR luminous high redshift sources.

We have presented scaling relations between 6.2 and 11.2$\mu$m PAH features and the [\ion{Ne}{2}] fine structure line against SFR ($\dot{M}_{\rm Sb}$) as well as component luminosities ($L_{\rm Sb}$, $L_{\rm Host}$, $L_{\rm AGN}^{o/c}$) from the radiative transfer modeling study of \citet{Efstathiou2022}. These fits provide a direct empirical calibration between $L_{\rm Sb}$ and $L_{\rm PAH}$ in the high SFR regime (50-2200 $\rm M_{\odot} yr^{-1}$).

For our sample of 42 ULIRGS, we find positive correlations between PAH emissions and, individually, starburst, host, total luminosity, and SFR. We explore the dependence of these relationships on properties of the starburst, AGN, and silicate depth, and find a significant influence from only the starburst age. We compare our results to those in other works \citep{Shipley2016,Maragkoudakis2018, Cortsen2019, Xie_Ho_2019,Mordini2021} and find differences over the luminosity range $L_{\rm Sb}=3.5\cdot 10^{11}-7.1\cdot 10^{12} \mathrm{L}_{\odot}$ of 0.3-0.9 dex for \citet{Cortsen2019}, 3.4-4.1 dex for \citet{Maragkoudakis2018}, 1.3-1.6 dex for \citet{Shipley2016}, 0.4-1.2 dex for \citet{Mordini2021}, and 0.5-1.4 dex for \citet{Xie_Ho_2019}. We employ methods from \citet{Kennicutt98} to verify that the differences between our work and others are not due to the variety of SFR calculation methods used. Our newly derived SFR relation differs from \citet{Shipley2016} by 1.1-1.3 dex, \citet{Mordini2021} by 0.4-0.9 dex, and \citet{Xie_Ho_2019} by 0.3-1 dex. 
We find negative correlations between PAH EW and AGN luminosity. These scaling relations are unaffected by the starburst age. 

We find positive correlations between [\ion{Ne}{2}] and starburst luminosity as well as SFR. As with the PAH features, there is a significant influence on the relationships from the starburst age. Our comparison with \citet{Ho_Keto_2007} shows more agreement, with an average difference of 0.2 dex. The correlation between [\ion{Ne}{3}] and starburst luminosity is weaker. We employ [\ion{Ne}{5}] as a tracer of AGN activity to investigate the AGN's effects on [\ion{Ne}{3}], suggesting that [\ion{Ne}{3}] originates from a combination of starburst and AGN ionization. 


Our conclusions are as follows:
\begin{itemize}
    \item Our results are consistent with PAH emission arising primarily in star-forming regions. The host galaxy contributes little to the observed PAH emission.
    \item Differences in our relations and those in previous studies \citep{Shipley2016, Maragkoudakis2018, Cortsen2019, Xie_Ho_2019, Mordini2021} can be explained by the differences between the samples explored in each work \citep[eg. ULIRGs in our work and Main Sequence SFGs][]{Cortsen2019}. These differences may reflect intrinsic differences in the starburst modes in ULIRGs and main sequence star formation at higher redshifts, suppression of PAH emissions due to AGN activity, or drowning of the PAHs by the AGN continuum.
    
    \item To account for these differences, and to avoid potentially underestimating the SFR by factors of up to 13, one should use the calibration in the proper luminosity range, derived from a parent sample most similar to the object in question. 
    
    \item Our results are consistent with AGN luminosity ``drowning out" PAH emission by producing a strong continuum, especially affecting the $6.2\mu$m feature. 
    
    \item Our results suggest that the 6.2$\mu$m and 11.2$\mu$m PAH features, and the [\ion{Ne}{2}] line, are better tracers of star formation than [\ion{Ne}{3}] in local ULIRGs. 
    
    \item We observe a positive correlation with starburst age for both the PAH relations and [\ion{Ne}{2}] relations. From our tests on this sample, we cannot distinguish whether this is due to a dependency on SFR or increased PAH/Ne heating efficiency in younger starbursts. 
\end{itemize}

These relations can be used to determine bolometric component luminosities of galaxies with MIR spectroscopy. This is especially relevant for studies of luminous starbursts at redshifts of $z < 3$ with JWST. Improved spectral resolution and sensitivity of the MIRI instrument allow for more accurate determinations of PAH luminosities \citep{Hernan-Caballero2020}, and (using these relations) more accurate component bolometric luminosities. 

Future work should focus on distinguishing the processes behind the observed dependence on starburst age and further investigating the possible quenching of Neon emission by the AGN. 

\begin{acknowledgments}
We thank the referee for a very helpful report. LR acknowledges support from the Research Experience for Undergraduate program at the Institute for Astronomy, University of Hawaii-Manoa funded through NSF grant \#2050710. LR would like to thank the Institute for Astronomy for their hospitality during the course of this project. EH acknowledges support from the Fundación Occident and the Instituto de Astrofísica de Canarias under the Visiting Researcher Programme 2022-2025 agreed between both institutions.
\end{acknowledgments}

\vspace{5mm}


\software{matplotlib \citep{Hunter:2007}, 
numpy \citep{harris2020array}, 
Scipy \citep{2020SciPy-NMeth}, 
pandas \citep{mckinney-proc-scipy-2010}}



\appendix 
\restartappendixnumbering
\section{}
\label{sec: appendix}

\begin{table}[ht]
\centering
\begin{tabular}{llllllllcl}
\hline
\hline
Name       & $L_{\rm Tot}$  & $L_{\rm 6.2}$        & $L_{\rm 11.2}$ & $L_{\rm NeII}$            & $L_{\rm NeIII}$  & 6.2 EW            & 11.2 EW           & Optical & Redshift     \\
           &  $[10^{12} \mathrm{L}_{\odot}]$             & $[10^9 \mathrm{L}_{\odot}]$ &                      & $[10^8 \mathrm{L}_{\odot}]$ &                  &                   &                   & Class   &       \\ \hline
00188-0856& $2.2^{+0.043}_{-0.05}$& $2.4^{+0.2}_{-0.2}$& $1.6^{+0.051}_{-0.051}$& $5.3^{+0.46}_{-0.46}$& $0.78^{+0.14}_{-0.14}$& $0.1^{+0.0049}_{-0.0049}$& $0.37^{+0.012}_{-0.012}$& 2.0& 0.128
\\
00397-1312& $9.4^{+0.34}_{-0.24}$& $1.3e+01^{+2.3}_{-2.3}$& $6.3^{+0.47}_{-0.47}$& $2.4e+01^{+1.6}_{-1.6}$& $1.5e+01^{+3.3}_{-3.3}$& $0.037^{+0.003}_{-0.003}$& $0.31^{+0.04}_{-0.04}$& 1.0& 0.262
\\
01003-2238& $1.6^{+0.12}_{-0.082}$& $1.3^{+0.1}_{-0.1}$& $0.93^{+0.049}_{-0.049}$& $3.0^{+0.3}_{-0.3}$& $1.2^{+0.13}_{-0.13}$& $0.055^{+0.0025}_{-0.0025}$& $0.041^{+0.001}_{-0.001}$& 1.0& 0.118
\\
03158+4227& $4.1^{+0.13}_{-0.15}$& $2.3^{+0.37}_{-0.37}$& $1.6^{+0.15}_{-0.15}$& $7.2^{+0.73}_{-0.73}$& $1.2^{+0.38}_{-0.38}$& $0.1^{+0.007}_{-0.007}$& $0.52^{+0.056}_{-0.056}$& 3.0& 0.134
\\
03521+0028& $2.4^{+0.029}_{-0.032}$& $2.5^{+0.16}_{-0.16}$& $1.9^{+0.082}_{-0.082}$& $4.6^{+0.21}_{-0.21}$& $2.1^{+0.13}_{-0.13}$& $0.35^{+0.011}_{-0.011}$& $0.73^{+0.034}_{-0.034}$& 2.0& 0.152
\\
05189-2524& $1.4^{+0.069}_{-0.047}$& $0.92^{+0.043}_{-0.043}$& $0.86^{+0.024}_{-0.024}$& $2.3^{+0.23}_{-0.23}$& $2.0^{+0.2}_{-0.2}$& $0.043^{+0.0009}_{-0.0009}$& $0.061^{+0.0014}_{-0.0014}$& 3.0& 0.043
\\
06035-7102& $1.6^{+0.03}_{-0.11}$& $2.3^{+0.15}_{-0.15}$& $1.4^{+0.036}_{-0.036}$& $2.8^{+0.22}_{-0.22}$& $0.71^{+0.14}_{-0.14}$& $0.11^{+0.0041}_{-0.0041}$& $0.25^{+0.007}_{-0.007}$& 1.0& 0.079
\\
06206-6315& $1.5^{+0.02}_{-0.028}$& $1.6^{+0.11}_{-0.11}$& $1.1^{+0.038}_{-0.038}$& $3.7^{+0.37}_{-0.37}$& $1.6^{+0.075}_{-0.075}$& $0.21^{+0.0078}_{-0.0078}$& $0.41^{+0.015}_{-0.015}$& 3.0& 0.092
\\
07598+6508& $4.2^{+0.16}_{-0.19}$& $2.0^{+0.2}_{-0.2}$& $1.2^{+0.096}_{-0.096}$& $6.1^{+0.062}_{-0.062}$& $3.8^{+0.78}_{-0.78}$& $0.011^{+0.0007}_{-0.0007}$& $0.015^{+0.0009}_{-0.0009}$& 4.0& 0.148
\\
08311-2459& $2.9^{+0.046}_{-0.083}$& $4.3^{+0.3}_{-0.3}$& $5.7^{+0.4}_{-0.4}$& $2.9^{+1.6}_{-1.6}$& $1.5e+01^{+0.74}_{-0.74}$& $0.14^{+0.006}_{-0.006}$& $0.2^{+0.0082}_{-0.0082}$& 4.0& 0.1
\\
08572+3915& $1.5^{+0.075}_{-0.053}$& $0.17^{+0.0}_{-0.0}$& $0.055^{+0.0}_{-0.0}$& $1.5^{+0.15}_{-0.15}$& $0.42^{+0.081}_{-0.081}$& $0.0031^{+0.0}_{-0.0}$& $0.0034^{+0.0}_{-0.0}$& 3.0& 0.058
\\
09022-3615& $1.7^{+0.023}_{-0.023}$& $2.8^{+0.13}_{-0.13}$& $2.0^{+0.047}_{-0.047}$& $1.3e+01^{+1.3}_{-1.3}$& $8.9^{+0.89}_{-0.89}$& $0.15^{+0.0038}_{-0.0038}$& $0.29^{+0.0082}_{-0.0082}$& 1.0& 0.06
\\
10378+1109& $1.7^{+0.041}_{-0.055}$& $1.2^{+0.14}_{-0.14}$& $0.78^{+0.049}_{-0.049}$& $4.5^{+0.45}_{-0.45}$& $0.77^{+0.2}_{-0.2}$& $0.1^{+0.007}_{-0.007}$& $0.36^{+0.017}_{-0.017}$& 2.0& 0.136
\\
10565+2448& $1.1^{+0.021}_{-0.0041}$& $2.3^{+0.089}_{-0.089}$& $1.7^{+0.028}_{-0.028}$& $7.3^{+0.73}_{-0.73}$& $0.86^{+0.086}_{-0.086}$& $0.64^{+0.018}_{-0.018}$& $0.61^{+0.012}_{-0.012}$& 1.0& 0.043
\\
11095-0238& $1.5^{+0.033}_{-0.035}$& $1.1^{+0.12}_{-0.12}$& $0.82^{+0.056}_{-0.056}$& $4.6^{+0.57}_{-0.57}$& $1.4^{+0.057}_{-0.057}$& $0.054^{+0.003}_{-0.003}$& $0.45^{+0.043}_{-0.043}$& 2.0& 0.107
\\
12071-0444& $2.2^{+0.06}_{-0.086}$& $2.7^{+0.18}_{-0.18}$& $1.2^{+0.073}_{-0.073}$& $5.9^{+0.31}_{-0.31}$& $5.8^{+0.19}_{-0.19}$& $0.12^{+0.0041}_{-0.0041}$& $0.076^{+0.0027}_{-0.0027}$& 3.0& 0.128
\\
13120-5453& $1.8^{+0.014}_{-0.015}$& $2.5^{+0.075}_{-0.075}$& $1.5^{+0.064}_{-0.064}$& $8.5^{+0.85}_{-0.85}$& $1.0^{+0.1}_{-0.1}$& $0.53^{+0.0051}_{-0.0051}$& $0.56^{+0.005}_{-0.005}$& 3.0& 0.031
\\
13451+1232& $1.9^{+0.19}_{-0.17}$& $0.23^{+0.1}_{-0.1}$& $0.36^{+0.065}_{-0.065}$& $5.1^{+0.62}_{-0.62}$& $5.2^{+0.53}_{-0.53}$& $0.011^{+0.0036}_{-0.0036}$& $0.014^{+0.0015}_{-0.0015}$& 3.0& 0.122
\\
14348-1447& $2.0^{+0.022}_{-0.087}$& $2.2^{+0.12}_{-0.12}$& $1.5^{+0.047}_{-0.047}$& $4.8^{+0.48}_{-0.48}$& $1.1^{+0.12}_{-0.12}$& $0.25^{+0.0055}_{-0.0055}$& $0.82^{+0.021}_{-0.021}$& 2.0& 0.083
\\
14378-3651& $1.1^{+0.012}_{-0.013}$& $1.5^{+0.091}_{-0.091}$& $0.93^{+0.029}_{-0.029}$& $3.3^{+0.33}_{-0.33}$& $0.35^{+0.035}_{-0.035}$& $0.44^{+0.014}_{-0.014}$& $0.53^{+0.012}_{-0.012}$& 3.0& 0.068
\\
15250+3609& $1.1^{+0.035}_{-0.034}$& $0.67^{+0.067}_{-0.067}$& $0.6^{+0.028}_{-0.028}$& $1.9^{+0.19}_{-0.19}$& $0.51^{+0.051}_{-0.051}$& $0.044^{+0.0021}_{-0.0021}$& $0.5^{+0.018}_{-0.018}$& 1.0& 0.055
\\
15462-0450& $1.3^{+0.13}_{-0.059}$& $1.5^{+0.1}_{-0.1}$& $1.0^{+0.042}_{-0.042}$& $4.9^{+0.49}_{-0.49}$& $0.91^{+0.17}_{-0.17}$& $0.079^{+0.0041}_{-0.0041}$& $0.088^{+0.0015}_{-0.0015}$& 4.0& 0.1
\\
16090-0139& $3.0^{+0.059}_{-0.066}$& $2.9^{+0.25}_{-0.25}$& $2.3^{+0.16}_{-0.16}$& $9.6^{+0.42}_{-0.42}$& $2.5^{+0.41}_{-0.41}$& $0.093^{+0.003}_{-0.003}$& $0.52^{+0.019}_{-0.019}$& 2.0& 0.134
\\
17208-0014& $2.2^{+0.027}_{-0.012}$& $2.5^{+0.062}_{-0.062}$& $1.6^{+0.035}_{-0.035}$& $4.6^{+0.45}_{-0.45}$& $0.91^{+0.1}_{-0.1}$& $0.46^{+0.0033}_{-0.0033}$& $1.2^{+0.027}_{-0.027}$& 1.0& 0.043
\\
19254-7245& $1.3^{+0.14}_{-0.036}$& $1.5^{+0.13}_{-0.13}$& $1.3^{+0.037}_{-0.037}$& $0.35^{+0.75}_{-0.75}$& $3.2^{+0.31}_{-0.31}$& $0.074^{+0.0035}_{-0.0035}$& $0.15^{+0.0041}_{-0.0041}$& 3.0& 0.062
\\
19297-0406& $2.5^{+0.083}_{-0.041}$& $3.1^{+0.17}_{-0.17}$& $2.0^{+0.043}_{-0.043}$& $8.4^{+0.083}_{-0.083}$& $1.2^{+0.12}_{-0.12}$& $0.44^{+0.011}_{-0.011}$& $0.85^{+0.016}_{-0.016}$& 1.0& 0.086
\\
20087-0308& $2.7^{+0.074}_{-0.052}$& $3.8^{+0.15}_{-0.15}$& $2.2^{+0.058}_{-0.058}$& $3.1^{+0.48}_{-0.48}$& $1.2^{+0.12}_{-0.12}$& $0.35^{+0.011}_{-0.011}$& $0.86^{+0.03}_{-0.03}$& 2.0& 0.106
\\
20100-4156& $3.2^{+0.095}_{-0.13}$& $3.4^{+0.24}_{-0.24}$& $2.6^{+0.11}_{-0.11}$& $8.4^{+1.1}_{-1.1}$& $3.2^{+0.73}_{-0.73}$& $0.14^{+0.0027}_{-0.0027}$& $0.91^{+0.041}_{-0.041}$& 1.0& 0.13
\\
20414-1651& $1.5^{+0.031}_{-0.032}$& $1.5^{+0.059}_{-0.059}$& $0.86^{+0.032}_{-0.032}$& $3.4^{+0.13}_{-0.13}$& $0.77^{+0.041}_{-0.041}$& $0.57^{+0.017}_{-0.017}$& $0.86^{+0.016}_{-0.016}$& 1.0& 0.087
\\
20551-4250& $1.0^{+0.051}_{-0.027}$& $1.0^{+0.081}_{-0.081}$& $0.71^{+0.029}_{-0.029}$& $1.5^{+0.15}_{-0.15}$& $0.31^{+0.032}_{-0.032}$& $0.11^{+0.0022}_{-0.0022}$& $0.33^{+0.0096}_{-0.0096}$& 1.0& 0.043
\\
22491-1808& $1.2^{+0.021}_{-0.032}$& $1.3^{+0.14}_{-0.14}$& $1.1^{+0.028}_{-0.028}$& $2.1^{+0.21}_{-0.21}$& $0.72^{+0.072}_{-0.072}$& $0.4^{+0.0094}_{-0.0094}$& $0.89^{+0.024}_{-0.024}$& 1.0& 0.078
\\
23128-5919& $0.74^{+0.018}_{-0.026}$& $1.4^{+0.036}_{-0.036}$& $1.1^{+0.063}_{-0.063}$& $3.3^{+0.034}_{-0.034}$& $2.5^{+0.25}_{-0.25}$& $0.36^{+0.008}_{-0.008}$& $0.3^{+0.0061}_{-0.0061}$& 1.0& 0.045
\\
23230-6926& $1.4^{+0.034}_{-0.039}$& $2.6^{+0.12}_{-0.12}$& $1.7^{+0.057}_{-0.057}$& $5.6^{+0.42}_{-0.42}$& $1.5^{+0.05}_{-0.05}$& $0.31^{+0.0094}_{-0.0094}$& $0.78^{+0.023}_{-0.023}$& 2.0& 0.107
\\
23253-5415& $2.0^{+0.097}_{-0.11}$& $1.4^{+0.067}_{-0.067}$& $1.3^{+0.06}_{-0.06}$& $6.3^{+0.34}_{-0.34}$& $2.2^{+0.32}_{-0.32}$& $0.29^{+0.0085}_{-0.0085}$& $0.48^{+0.015}_{-0.015}$& 2.0& 0.13
\\
23365+3604& $1.4^{+0.044}_{-0.028}$& $1.5^{+0.093}_{-0.093}$& $1.1^{+0.043}_{-0.043}$& $2.2^{+0.22}_{-0.22}$& $0.19^{+0.026}_{-0.026}$& $0.4^{+0.014}_{-0.014}$& $0.48^{+0.018}_{-0.018}$& 2.0& 0.064
\\
09320+6134& $1.0^{+0.034}_{-0.013}$& $1.6^{+0.093}_{-0.093}$& $1.0^{+0.024}_{-0.024}$& $3.2^{+0.32}_{-0.32}$& $1.3^{+0.13}_{-0.13}$& $0.25^{+0.0098}_{-0.0098}$& $0.51^{+0.013}_{-0.013}$& 2.0& 0.039
\\
12540+5708& $3.5^{+0.21}_{-0.14}$& $1.4^{+0.15}_{-0.15}$& $1.2^{+0.06}_{-0.06}$& $1.1^{+0.21}_{-0.21}$& $0.33^{+0.043}_{-0.043}$& $0.021^{+0.0006}_{-0.0006}$& $0.035^{+0.0008}_{-0.0008}$& 4.0& 0.042
\\
13428+5608& $1.3^{+0.041}_{-0.027}$& $1.3^{+0.079}_{-0.079}$& $0.68^{+0.017}_{-0.017}$& $3.4^{+0.34}_{-0.34}$& $2.7^{+0.27}_{-0.27}$& $0.19^{+0.0042}_{-0.0042}$& $0.36^{+0.0051}_{-0.0051}$& 3.0& 0.037
\\
13536+1836& $0.76^{+0.063}_{-0.048}$& $0.079^{+0.0}_{-0.0}$& $0.38^{+0.07}_{-0.07}$& $1.4^{+0.16}_{-0.16}$& $6.1^{+0.61}_{-0.61}$& $0.0024^{+0.0}_{-0.0}$& $0.022^{+0.0023}_{-0.0023}$& 3.0& 0.049
\\
15327+2340& $1.1^{+0.066}_{-0.0069}$& $0.48^{+0.012}_{-0.012}$& $0.33^{+0.015}_{-0.015}$& $1.3^{+0.13}_{-0.13}$& $0.16^{+0.016}_{-0.016}$& $0.27^{+0.0021}_{-0.0021}$& $0.72^{+0.013}_{-0.013}$& 3.0& 0.018
\\
16504+0228& $0.68^{+0.029}_{-0.032}$& $1.3^{+0.047}_{-0.047}$& $1.3^{+0.029}_{-0.029}$& $6.1^{+0.61}_{-0.61}$& $2.2^{+0.22}_{-0.22}$& $0.38^{+0.0033}_{-0.0033}$& $0.72^{+0.011}_{-0.011}$& 2.0& 0.024
\\
01572+0009& $3.4^{+0.11}_{-0.088}$& $3.0^{+0.28}_{-0.28}$& $3.2^{+0.17}_{-0.17}$& $1.3e+01^{+1.3}_{-1.3}$& $1.9e+01^{+1.9}_{-1.9}$& $0.064^{+0.0036}_{-0.0036}$& $0.067^{+0.0017}_{-0.0017}$& 4.0& 0.163
\\ \hline
\end{tabular}
\caption{Table of PAH and Neon luminosities, as well as PAH EW and Redshift for the sample used in this study. For more information on this sample, see tables 1 and 4 in \citet{Efstathiou2022}. Three objects, IRAS 23128-5919, IRAS 13536+1836, and IRAS 16504+0228, were identified as ULIRGs under an Einstein-DeSitter cosmological model with \mbox{$H_0 = 50$\,km\,s$^{-1}$\,Mpc$^{-1}$}. Under a $\Lambda \rm CDM$ cosmological model, these objects' total luminosity falls below the $10^{12}\rm L_{\odot}$ boundary, but we keep them in our sample for completeness. }
\label{table: sample}
\end{table}
\newpage

\begin{figure*}[ht]
\includegraphics[width=16cm]{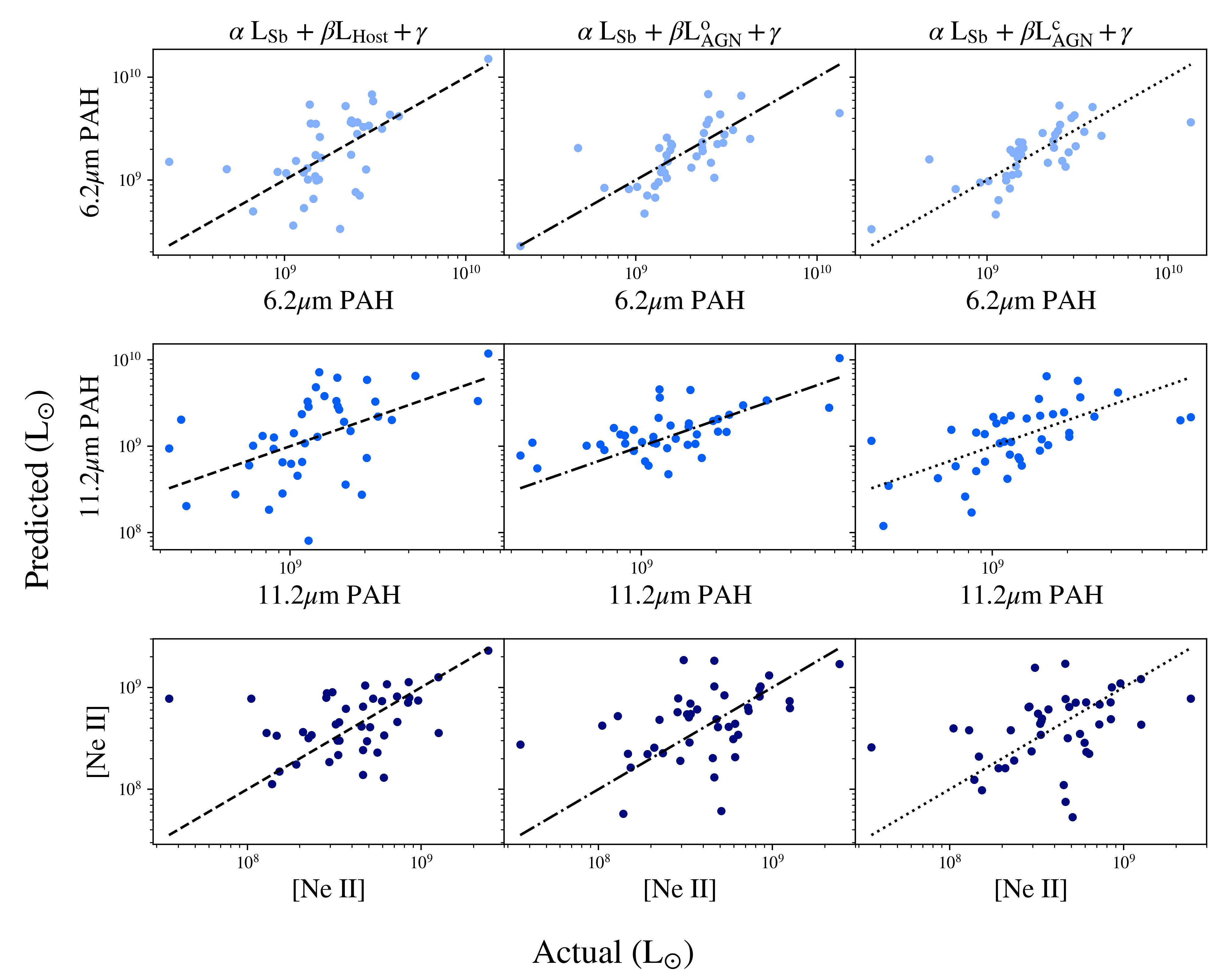}
\centering
\caption{Plots detailing multi-variable Relations. Plots show predicted luminosity versus actual observed values, with lines of slope unity plotted atop the data. The form of each relation is shown at the top of the column, while the corresponding line luminosity is shown at the left end of each row, and along the x-axis of each plot.}
\label{fig: multivar}
\end{figure*}

\begin{table}[ht]
\begin{tabular}{|l|l|l|}
\hline
               & $L_{\rm 6.2}$ & $L_{\rm 11.2}$ \\ \hline
$L_{\rm Sb}$       &  0.62, $1.7 \cdot 10^{-8}$&  0.59, $6.17 \cdot 10^{-8}$\\ \hline
$\dot{M}_{\rm Sb}$  &     0.65, $2.8 \cdot 10^{-9}$&   0.61, $1.95 \cdot 10^{-8}$\\ \hline
$L_{\rm Host}$ &     0.13, 0.24&    0.19, 0.09\\ \hline
$L^{o}_{\rm Tot}$ &       0.44,   $6.76 \cdot 10^{-5}$&    0.48, $1.07 \cdot 10^{-5}$\\\hline
 $L^{o}_{\rm AGN}$& $-$0.03, 0.80&0.01, 0.90\\\hline
 $L^{c}_{\rm AGN}$& $-$0.14, 0.19&$-$0.04, 0.70\\\hline 
$\tau_{\rm UV}$    &  $-$0.32, $3.3 \cdot 10^{-3}$&    $-$0.22,      0.04\\ \hline
$L_{\rm Ne II}$    &  0.42, $1.6 \cdot 10^{-4}$&    0.44, $5.26\cdot 10^{-5}$\\ \hline
$L_{\rm Ne III}$    &  0.31, $5.6 \cdot 10^{-3}$&    0.25,      0.02\\ \hline
\end{tabular}
\quad
\begin{tabular}{|l|l|l|}
\hline
              & $6.2 \mu$m EW & $11.2 \mu$m EW \\ \hline
$L^{o}_{\rm AGN}$ &   $-$0.54,    $8.80 \cdot 10^{-7}$&      $-$0.36, $1.04 \cdot 10^{-3}$\\ \hline
$L^{c}_{\rm AGN}$ &   $-$0.35, $1.5 \cdot 10^{-3}$&       $-$0.12, 0.25\\ \hline
$\tau_{\rm Sb}$         &   $-$0.39, $3.6 \cdot 10^{-4}$&      $-$0.38, $4.58 \cdot 10^{-4}$\\ \hline
$SB_{\rm age}$         &  0.46,  $2.60 \cdot 10^{-5}$&    0.41, $1.40 \cdot 10^{-4}$\\ \hline
$SB_{tv}$          &   0.40, $3.0 \cdot 10^{-4}$&     0.49, $7.03 \cdot 10^{-6}$\\ \hline
SiDep         &     NA           &    0.35, $1.37 \cdot 10^{-3}$\\ \hline
$\theta_{inc}$     &   NA             &     0.36, $8.85 \cdot 10^{-4}$\\ \hline
\end{tabular}
\caption{Summary of $\tau$ and P-values for Kendall-$\tau$ correlation tests between  $ 6.2 \mu$m and  $11.2\mu$m luminosities ($L_{\rm 6.2}, L_{\rm 11.2}$; left), and 6.2$\mu$m and 11.2$\mu$m EW (right) against other variables. Cells are of the form ($\tau$, p).}
\label{Table: pahkt}
\end{table}

\begin{table}[ht]
\centering
\begin{tabular}{|l|l|l|l|}
\hline
        & $L_{\rm Ne II}$        & $L_{\rm Ne III}$            &$L_{\rm Ne II} + L_{\rm Ne III}$\\ \hline
$L_{\rm 6.2}$   & 0.42, $1.6 \cdot 10^{-4}$& 0.31, $5.6 \cdot 10^{-3}$&0.47, $2.0 \cdot 10^{-5}$\\ \hline
$L_{\rm 11.2}$   & 0.44, $5.26 \cdot 10^{-5}$& 0.25, 0.02&0.40, $1.7 \cdot 10^{-4}$\\ \hline
$L_{\rm Sb}$     & 0.38, $3.9\cdot 10^{-4}$  & 0.15, 0.16&0.34, $1.5 \cdot 10^{-3}$\\ \hline
$L^{o}_{\rm Tot}$    & 0.37, $5.9\cdot 10^{-4}$  & 0.24, $2.6 \cdot 10^{-2}$   &0.37, $5.4 \cdot 10^{-4}$\\ \hline
$\dot{M}_{\rm Sb}$     & 0.35, $9.5\cdot 10^{-4}$  & 0.12, 0.27                          &0.29, $7.4 \cdot 10^{-3}$\\\hline
 $L^{o}_{\rm AGN}$
& 0.08, 0.48&0.20, 0.07 &0.17, 0.12\\\hline
 $L^{c}_{\rm AGN}$& $-$0.02, 0.87&0.05, 0.66 &0.02, 0.9\\\hline 
$11.2 \mu$m EW  & NA                        & $-$0.21, 0.04&NA\\ \hline
$\tau_{\rm Sb}$   & NA                        & 0.24, $2.5 \cdot 10^{-2}$   &NA\\ \hline
SiDep   & NA                        & 0.25, $1.8 \cdot 10^{-2}$&NA\\ \hline
$SB_{tv}$    & NA                        & $-$0.33, $2.3 \cdot 10^{-3}$  &NA\\ \hline
\end{tabular}
\caption{Summary of $\tau$ and P-values for Kendall-$\tau$ correlation tests between [\ion{Ne}{2}] (12.8$\mu$m) and [\ion{Ne}{3}] (15$\mu$m) luminosities ($L_{\rm Ne II}, L_{\rm Ne III}$) and other variables. Cells are of the form ($\tau$, p). }
\label{table: nekt}
\end{table}

\begin{table}[]
\centering
\begin{tabular}{l|l|l|l|}
\cline{2-4}
                                                       & $L_{\rm 6.2}$                   & $L_{\rm 11.2}$                  & $L_{\rm Ne II }$          \\ \hline
\multicolumn{1}{|l|}{$L_{\rm Sb}$}                     & $1.09 \pm 0.16$, $-3.88 \pm 1.92$& 
$0.87 \pm 0.14$, $-1.38 \pm 1.64$& $0.67 \pm 0.09$, $0.56 \pm 1.06$\\ \hline
\multicolumn{1}{|l|}{$L_{\rm Sb + Host}$}                  & $1.19 \pm 0.17$, $-5.22 \pm 2.05$& $0.99 \pm 0.14$, $-2.85 \pm 1.76$& $0.73\pm 0.09$, $-0.16 \pm 1.13$                                    \\ \hline
\multicolumn{1}{|l|}{$\dot{M}_{\rm Sb}$}               & $0.95 \pm 0.12$, $6.81 \pm 0.31$& $1.08 \pm 0.16$, $6.32 \pm 0.42$& $0.92 \pm 0.16$, $6.24 \pm 0.42$\\ \hline
\multicolumn{1}{|l|}{$L_{\rm Sb} + L_{\rm Host}$}      & $0.87 \pm 0.23$, $0.74 \pm 0.20$, $-9.26 \pm 3.56$& $0.74 \pm 0.29$, $1.09 \pm 0.40$, $-11.68 \pm 5.24$& $0.65 \pm 0.18$, $0.56 \pm 0.16$, $-5.28 \pm 2.57$                                    \\ \hline
\multicolumn{1}{|l|}{$L_{\rm Ne II}$}                  & $0.90 \pm 0.10$, $1.47 \pm 0.90$& $0.88 \pm 0.088$, $1.45 \pm 0.77$& NA                                    \\ \hline
\multicolumn{1}{|l|}{$L_{\rm Sb}$ + $L^{o}_{\rm AGN}$} & $1.05 \pm 0.17$, $-0.48 \pm 0.11$, $2.03 \pm 1.74$& $1.33 \pm 0.27$, $-0.74 \pm 0.21$, $1.52 \pm 2.47$& $1.16 \pm 0.20$, $-0.42 \pm 0.14$, $-0.47 \pm 1.89$\\ \hline
\multicolumn{1}{|l|}{$L_{\rm Sb}$ + $L^{c}_{\rm AGN}$} & $0.85 \pm 0.13 \space $, $-0.30 \pm 0.07$, $2.51 \pm 1.43$& $1.05 \pm 0.20$, $-0.50 \pm 0.14$, $2.30 \pm 2.17$& $0.97 \pm 0.16$, $-0.41 \pm 0.11$, $1.87 \pm 1.76$\\ \hline
\end{tabular}
\caption{Table of scaling relation parameters $\alpha$ and $\beta$, with uncertainties, for  $ L_{\rm 6.2}$ and  $ L_{\rm 11.2}$,  and $L_{\rm Ne II }$ against other variables. Cells are of the form ($\alpha \pm \sigma_{\alpha}$, $\beta \pm \sigma_{\beta}$,  $\gamma \pm \sigma_{\gamma}$) corresponding to $ \mathcal{L}_{\rm Observable}$ = $(\alpha \pm \sigma_{\alpha}) \times \mathcal{L}_{\rm Component}$ + $(\beta \pm \sigma_{\beta})$ for single variable relations and $ \mathcal{L}_{\rm Observable}$ = $(\alpha \pm \sigma_{\alpha}) \times \mathcal{L}_{\rm Component A} + (\beta \pm \sigma_{\beta}) \times \mathcal{L}_{\rm Component B} + (\gamma \pm \sigma_\gamma)$ for multi-variable relations.}
\label{table: pah_scaling_relations}
\end{table}

\begin{table}[ht]
\centering
\begin{tabular}{|l|l|l|}
\hline
                       & $6.2 \mu$m EW                         & $11.2\mu$m EW                          \\ \hline
$ L^o_{\rm AGN}$           & $-$1.42 $\pm$ 0.15, 15.51 $\pm$ 1.68& $-$1.95 $\pm$ 0.33, 21.91 $\pm$ 3.83\\ \hline
$ L^c_{\rm AGN}$           & $-$1.96 $\pm$ 0.47, 22.57 $\pm$ 5.66& NA                                     \\ \hline
$L_{\rm Ne III (15\mu m)}$ & NA                                     & $-$1.62 $\pm$ 0.35, 12.93 $\pm$ 2.93\\ \hline
\end{tabular}
\label{table: EW_scaling_relations}
\caption{Table of scaling relation parameters $\alpha$ and $\beta$, with uncertainties, for  $6.2 \mu$m EW, $11.2\mu$m EW against other variables. Cells are of the form ($\alpha \pm \sigma_{\alpha}$, $\beta \pm \sigma_{\beta}$) corresponding to $ \mathcal{W}_{\rm Observable}$ = $(\alpha \pm \sigma_{\alpha}) \times \mathcal{L}_{\rm Component}$ + $(\beta \pm \sigma_{\beta})$.}
\end{table}

\begin{table}[]
\centering
\begin{tabular}{|l|l|l|}
\hline
                                     & Low                           & High                          \\ \hline
$SB_{\rm age}$ (Myr)                     & $5.2-31.8$& $31.8-34.9$                     \\ \hline
$\tau_{\rm Sb}$                          & $(1.5 - 1.9) \cdot 10^7$ & $(1.9-3.5) \cdot 10^7$ \\ \hline
$SB_{tv}$                            & $51.1-169.8$                    & $169.8-248.9$                   \\ \hline
$\dot{M}_{\rm Sb}$ ($\mathrm{M}_{\odot} \mathrm{yr}^{-1}$) & $87.1-373.6$& $373.6- 2192.0$                 \\ \hline
$L_{\rm Sb}/L_{\rm Tot}$             & $0.38-0.76$                     & $0.76-0.94$                     \\ \hline
\end{tabular}
\caption{Table of parameter ranges when the sample is divided into two at the median starburst age.}
\end{table}

\newpage
\clearpage
\bibliography{citation}{}
\bibliographystyle{aasjournal}



\end{document}